\documentclass[a4paper,11pt]{article}
\pdfoutput=1 

\usepackage{jcappub}
\usepackage[T1]{fontenc}
\usepackage{multirow}
\usepackage{url}
\usepackage{graphicx}
\usepackage{subfigure}
\usepackage{amsmath}
\usepackage{enumitem}
\usepackage{braket}
\usepackage{MnSymbol}
\usepackage{comment}
\usepackage{nicematrix}
\NiceMatrixOptions
  {
    code-for-first-row = \color{gray} ,
    code-for-last-col = \color{gray} ,
    xdots/horizontal-labels
  }
\usepackage{scalerel}
\newlength\bshft
\def\fakebold#1{\ThisStyle{\ooalign{$\SavedStyle#1$\cr%
  \kern-\bshft$\SavedStyle#1$\cr%
  \kern\bshft$\SavedStyle#1$}}}

\DeclareMathAlphabet{\mathbbmsl}{U}{bbm}{m}{sl}

\def\healpix{\texttt{HEALPix}}

\def\nside{\textit{N}_{\mathrm{side}}}

\def\estimated{estimated}
\def\cmbrealizations{200}

\newcommand{\ratio}[0]{tensor-to-scalar ratio}
\newcommand{\quotes}[1]{``#1''}

\title{\boldmath Mitigating half-wave plate systematics at the map-making level: calibration requirements for \textit{LiteBIRD}}

\author[1]{N.\,Raffuzzi,}
\author[2,3]{A.\,Carones,}
\author[4]{M.\,Monelli,}
\author[5]{S.\,Giardiello,}
\author[1,6,7]{L.\,Pagano,}
\author[8,4]{Y.\,Sakurai,}
\author[9]{H.\,Ishino,}
\author[10]{E.\,Allys,}
\author[11]{A.\,Anand,}
\author[12]{J.\,Aumont,}
% ...
\affiliation[1]{Dipartimento di Fisica e Scienze della Terra, Università di Ferrara, Via Saragat 1, 44122 Ferrara, Italy}
\affiliation[2]{International School for Advanced Studies (SISSA), Via Bonomea 265, 34136, Trieste, Italy}
\affiliation[3]{INFN Sezione di Trieste, via Valerio 2, 34127 Trieste, Italy}
\affiliation[4]{Kavli Institute for the Physics and Mathematics of the Universe (Kavli IPMU, WPI), UTIAS, The University of Tokyo, Kashiwa, Chiba 277-8583, Japan}
\affiliation[5]{School of Physics and Astronomy, Cardiff University, Cardiff CF24 3AA, UK}
\affiliation[6]{INFN Sezione di Ferrara, Via Saragat 1, 44122 Ferrara, Italy}
\affiliation[7]{Université Paris-Saclay, CNRS, Institut d’Astrophysique Spatiale, 91405, Orsay, France}
\affiliation[8]{Suwa University of Science, Chino, Nagano 391-0292, Japan}
\affiliation[9]{Okayama University, Department of Physics, Okayama 700-8530, Japan}
\affiliation[10]{Laboratoire de Physique de l’École Normale Supérieure, ENS, Université PSL, CNRS, Sorbonne Université, Université de Paris, 75005 Paris, France}
\affiliation[11]{Dipartimento di Fisica, Università di Roma Tor Vergata, Via della Ricerca Scientifica, 1, 00133, Roma, Italy}
\affiliation[12]{IRAP, Université de Toulouse, CNRS, CNES, UPS, Toulouse, France}
% ...

\emailAdd{nicolelia.raffuzzi@unife.it}
\emailAdd{acarones@sissa.it}
\emailAdd{marta.monelli@ipmu.jp}

\abstract{Although half-wave plates (HWPs) are becoming a popular choice of polarization modulators for cosmic microwave background (CMB) experiments, their non-idealities can introduce systematic effects that should be carefully characterized and mitigated. One possible mitigation strategy is to incorporate information about the non-idealities at the map-making level, which helps to reduce the HWP-induced distortions of the reconstructed CMB. Nevertheless, the non-idealities can only be known with finite precision. In this paper we investigate the consequences of discrepancies between their true frequency profiles and those assumed by the map-maker. We present an end-to-end framework, including a blind component-separation step, and use it to translate these discrepancies into a bias on the tensor-to-scalar ratio, $r$, for the \emph{LiteBIRD} satellite mission. We subsequently derive realistic and conservative measurement requirements for accurately characterizing the HWP non-idealities to ensure they do not compromise \emph{LiteBIRD}'s ambitious scientific goals. We find that the obtained results are robust against sky models with varying complexity.}

\begin{document}

\maketitle

\section{Introduction} \label{sec:intro}
Observations of the cosmic microwave background (CMB) temperature and polarization anisotropies offer a unique window into the early Universe, providing insights into its fundamental properties and structure. While current observations have nearly exhausted the physical information encoded in temperature anisotropies~\cite{Planck:2018vyg, WMAP:2012fli, ACT:2020frw, SPT-3G:2025bzu}, the nearer polarized signal, in particular $B$ modes, is in early stages of exploration~\cite{SPT-3G:2025bzu, BICEP:2021xfz, AtacamaCosmologyTelescope:2025vnj} and represents the primary focus of the next generation of CMB experiments~\cite{LiteBIRD:2022cnt, SimonsObservatory:2018koc}.

According to the standard cosmological model, the early Universe underwent a period of accelerated expansion, known as cosmic inflation~\cite{Starobinsky:1980te, Guth:1980zm, Linde:1981mu, Linde:1983gd, Mukhanov:1981xt}. During this phase, quantum fluctuations were stretched from microscopic to cosmological scales, giving rise to both scalar and tensor perturbations. The relative amplitude of these perturbations is commonly quantified by the \ratio\ $r$, defined as the ratio of tensor to scalar perturbation amplitudes. Tensor perturbations imprint a distinctive curl-like $B$-mode polarization pattern in the CMB~\cite{Zaldarriaga:1996xe, Seljak:1996gy, Kamionkowski:1996ks} and detecting this signal would provide a direct probe of primordial gravitational waves and a powerful test of the inflationary paradigm. The expected amplitude of this signal is however much smaller than that of temperature anisotropies, making it extremely challenging to detect.
The most stringent current upper limits yield $r \leq 0.032$ (95\% C.L.)~\cite{Tristram:2021tvh}, with future missions targeting a first detection with an uncertainty at $\sim 10^{-3}$.

The \textit{LiteBIRD} satellite mission is one of the next-generation experiments designed to measure large-scale CMB polarization with unprecedented sensitivity. Scheduled to be launched in the $2030$s, \textit{LiteBIRD} will observe the entire sky across a wide frequency range, from $34$ to $440$ GHz~\cite{LiteBIRD:2022cnt}, with the primary objective of detecting the $B$-mode polarization signal on large angular scales. These are the scales where primordial gravitational waves leave their most significant imprint, with key spectral features such as the reionization bump ($\ell \lesssim 10$) and the recombination bump ($\ell \sim 80$)~\cite{LiteBIRD:2020khw}. Achieving a detection of primordial $B$ modes would allow \textit{LiteBIRD} to constrain the energy scale of inflationary models by measuring $r$ with a target sensitivity of $\sigma_r \leq 10^{-3}$. 
The mission's precise large-scale measurements and robust foreground control will provide essential all-sky data, which, combined with deep ground-based surveys (e.g. BICEP/Keck~\cite{BICEP:2021xfz}, SPT-3G D1~\cite{SPT-3G:2025bzu}, ACT DR6~\cite{AtacamaCosmologyTelescope:2025blo}, Simons Observatory~\cite{SimonsObservatory:2018koc}), will significantly advance our understanding of the early Universe.

Foreground contamination from the Milky Way presents one of the most significant challenges \emph{LiteBIRD} will face in detecting $B$ modes~\cite{Planck:2015mvg}. Galactic synchrotron and thermal dust emission are the dominant polarized signals at low and high frequencies, respectively. While emission processes like free–free radiation, anomalous microwave emission (AME), and CO molecular lines can generally be neglected in polarization studies due to their low polarization fractions, synchrotron and thermal dust signals must be carefully modeled and removed. To this end, \textit{LiteBIRD} will rely on advanced component separation techniques to isolate the CMB signal from the foregrounds, made possible by the large frequency coverage. Two general approaches are typically used: parametric-fitting methods, which explicitly model foreground components (with physical parameters of foregrounds and their frequency dependence) and marginalize over them during likelihood analysis~\cite{Stompor:2008sf}; and blind methods, like the Needlet Internal Linear Combination (NILC)~\cite{Delabrouille:2008qd,Carones:2022xzs}, which reconstruct the CMB as the minimum-variance solution from the linear combination of data from multiple frequency bands. Parametric fitting methods allow for better separation in complex regions at the price of requiring accurate modeling of component spectral energy distributions (SEDs), while blind methods, without assumptions about the SEDs, offer robustness when foreground properties are more nuanced.

Systematic effects pose another significant challenge for CMB polarization experiments, and thoughtful design choices are essential to mitigate them. For instance, \textit{LiteBIRD} will use a rotating half-wave plate (HWP) as a polarization modulator, which will help suppress instrumental $1/f$ noise and reduce pair-differencing systematics~\cite{Johnson:2006jk}. However, realistic HWPs are inevitably characterized by non-idealities that can result in additional systematic effects, which, if not accounted for in the analysis, can compromise measurement accuracy. For instance, the response of any realistic HWP is inherently frequency dependent, which can introduce biases in the recovered polarization maps and power spectra. One way to address this issue is to incorporate information about the non-idealities at the map-making level~\cite{Giardiello:2021uxq}, which can help mitigate HWP-induced distortions in the reconstructed CMB. The more accurately the HWP specifics are known, the more effective the mitigation.

In this work, we focus on the inevitable mismatch between the true and estimated frequency-dependent HWP non-idealities, and assess its impact on the inference of the tensor-to-scalar ratio, $r$. Specifically, we consider only the spectral dependence of the HWP Mueller matrix, neglecting potential variations with the incidence angle of incoming light. This choice reflects the focus of the current analysis on band-averaged effects. For treatments incorporating incidence angle dependence, see Ref.~\cite{Patanchon:2023ptm}. We also do not model effects such as emissivity due to temperature variations and its coupling to detector nonlinearities, HWP wobble, or wedge-like geometric systematics that may couple to pointing inaccuracies. These simplifications set the range of validity of our analysis, which primarily characterizes biases arising from (time-stationary) spectral mismatches under ideal optical alignment assumptions. To do so, we simulate the maps reconstructed in the presence of a mismatch, propagate the distortions through a realistic minimum-variance component-separation procedure, and infer the resulting bias on $r$. We adopt the \textit{LiteBIRD} specifications~\cite{LiteBIRD:2022cnt} as instrumental settings for the analysis. Using this framework, we derive requirements on the accuracy of HWP calibration to ensure that the resulting systematic bias, due to the HWP distortion, remains within the budget  defined by the \textit{LiteBIRD} Collaboration~\cite{LiteBIRD:2022cnt}.

The structure of this paper is as follows. Section~\ref{sec: methodology} outlines the methodology used throughout this work. In section~\ref{subsec: sims}, we describe how the HWP non-idealities are incorporated at the scanning and map-making levels to generate simulations. Section~\ref{subsec: comp sep} then details the component-separation algorithm used to recover the CMB signal, followed by the computation of the power spectrum and likelihood maximization in section~\ref{subsec: spectra and likelihood} to assess the resulting impact on $r$. The procedure to derive requirements on the miscalibration of HWP parameters is outlined in section~\ref{subsec: reqs}. The results of this analysis are presented in section~\ref{sec: results}, and we draw conclusions in section~\ref{sec: conclusions}. Appendices~\ref{sec: appendix effective mueller} and~\ref{sec: appendix validation of map-based} provide detailed derivations and validation of the simulations used throughout this work.
\section{Methodology} \label{sec: methodology}
To investigate how a miscalibration of the \estimated\ HWP parameters affects the bias on the \ratio, we follow a structured approach. First, we simulate a set of band-integrated maps for each \textit{LiteBIRD} frequency channel, each comprising \cmbrealizations\ realizations of CMB, noise, and specified systematic effects. Next, we apply a blind component separation technique to recover a cleaned CMB $B$-mode signal, estimate the angular power spectra, and derive $r$. This procedure is repeated for each combination of sky model and HWP parameters, as outlined in detail in section~\ref{subsec: reqs}. In the following, we examine each of these steps in turn.

\subsection{Simulating the band-integrated maps} \label{subsec: sims}
\subsubsection{Mathematical framework}
To capture the impact of HWP systematics in the \textit{LiteBIRD} data processing pipeline, we incorporate the HWP properties directly into the map-making procedure~\cite{Giardiello:2021uxq}, which, for the scope of this study, is a simple binning algorithm. In CMB experiments, the raw time-ordered data (TOD) collected by detectors is converted into pixelized sky maps through the map-making:
\begin{align} \label{eq:cartoon map2tod_tod2map}
    \mathbf{d} &= A_\text{true} \; \mathbf{s}_\text{in}\,, \\ 
    \mathbf{s}_\text{out} &= (A_\text{est}^\intercal A_\text{est})^{-1} A_\text{est}^\intercal \mathbf{d}\,.
\end{align}
Here, $\mathbf{d}$ represents the TOD, with dimensions given by the number of time samples, $n_\text{samples}$\footnote{For simplicity, in these formulae we consider an idealized, noiseless case with a single detector. Extending this to multiple detectors involves adding an additional dimension for the total number of detectors, making the TOD two-dimensional, with dimensions $n_\text{det} \times n_\text{samples}$, and the response matrix three-dimensional, with dimensions $n_\text{det} \times n_\text{samples} \times n_\text{pix}$.}; $\mathbf{s}_\text{in/out}$ denotes the input and output pixelized sky maps, whose dimensions are determined by the number of pixels at the chosen resolution; $A_{\text{true}}$ is the true response matrix of the instrument during the scanning of the sky, while $A_{\text{est}}$ is the estimated response matrix used in the map-making process. Both $A_{\text{true}}$ and $A_{\text{est}}$ have dimensions $n_\text{samples} \times n_\text{pix}$ for a single detector case. 

The overall map-making kernel, which links the reconstructed sky signal with the true one, is given by $(A_{\text{est}}^\intercal A_{\text{est}})^{-1} A_{\text{est}}^\intercal  A_{\text{true}}$. Therefore, any difference between the two response matrices induces deviations of $\mathbf{s}_\text{out}$ from $\mathbf{s}_\text{in}$. Both $A_{\text{true}}$ and $A_{\text{est}}$ are sparse matrices with non-zero entries indicating which pixels are observed at each time sample. The non-zero elements of the two matrices are respectively characterized by the true and estimated optical chains, $\mathbb{S}_{\text{true}}$ and $\mathbb{S}_{\text{est}}$. In this analysis, we express the optical chain as
\begin{equation} \label{eq:optical_chain}
    \mathbb{S} = \mathbf{a}^\intercal  \mathcal{M}_{\text{pol}} \mathcal{R}_{\xi-\theta} \mathcal{M}_\textsc{hwp} \mathcal{R}_{\theta+\psi}, \quad \text{ with } \quad \mathbf{a}^\intercal =\left(\begin{array}{lll}
1 & 0 & 0
\end{array}\right),
\end{equation}
where $\mathcal{R}_{\xi-\theta}$ and $\mathcal{R}_{\theta+\psi}$ are rotation matrices describing the relative tilt between the spinning HWP reference frame and, respectively, the sky coordinates and the focal plane, while $\mathcal{M}_{\text{pol}}$ and $\mathcal{M}_{\textsc{hwp}}$ are Mueller matrices accounting for, respectively, the polarimeter orientation and the HWP in its rest frame. For more details, we refer the interested readers to Ref.~\cite{Monelli:2022pru}. We assume the HWP Mueller matrix to exhibit a frequency dependence:
\begin{equation}\label{eqn:HWPmueller}
    \mathcal{M}_\textsc{hwp}(\nu) =
 \begin{pmatrix}
  m_{II}(\nu) & m_{IQ}(\nu) & m_{IU}(\nu) \\
  m_{QI}(\nu) & m_{QQ}(\nu) & m_{QU}(\nu) \\
  m_{UI}(\nu) & m_{UQ}(\nu) & m_{UU}(\nu)
 \end{pmatrix}.
\end{equation}
where $I$ is the Stokes parameter for total intensity and $Q, U$ are the linear polarization modes.
Here, we omit terms associated with circular polarization, since $V$ modes are expected to be negligible within the standard cosmological model because they are not generated by Thomson scattering during recombination or reionization. However, other studies have relaxed this assumption to explore potential production mechanisms and detection strategies for $V$ modes~\cite{Raffuzzi:2024wyh}.

In this study, we focus on the impact of discrepancies between the response matrices $A_{\text{true}}$ and $A_{\text{est}}$ (i.e., between the optical chains $\mathbb{S}_{\text{true}}$ and $\mathbb{S}_{\text{est}}$) induced by differences due to miscalibration between the true and the estimated HWP Mueller matrices ($\mathcal{M}_{\textsc{hwp}}^{\text{est}}(\nu) \neq \mathcal{M}_{\textsc{hwp}}^{\text{true}}(\nu)$).
As a result of these discrepancies, the signal at a given frequency channel $i$, $\mathbf{s}^i_\text{out}$, reconstructed from the TOD through map-making, can be obtained from the true sky maps, $\mathbf{s}_\text{in}(\nu)$, considering an effective Mueller matrix $\mathcal{M}_\text{eff}^i(\nu)$, where we are assuming perfect cross-linking\footnote{By \quotes{perfect cross-linking} we mean that the linear combinations of angles $\theta+\psi$ and $\xi-\theta$ are sampled uniformly enough that all orientations and HWP angles are equally represented in each pixel. With LiteBIRD’s scan strategy and rapid HWP rotation, this is expected to hold to good approximation.}, i.e., averaging over all angles for a single pixel (see Appendix~\ref{sec: appendix effective mueller}):
\begin{equation}\label{eqn:band_int_out_map_wMeff}
    \mathbf{s}^i_\text{out} \!\simeq\! \sum_\lambda\! \int_{\nu^i_\text{min}}^{\nu^i_\text{max}}
    \! \frac{\text{d}\nu}{\Delta\nu^i}\, \mathcal{M}_\text{eff}^i(\nu)\, \mathbf{s}^{\,i}_{\mathrm{in},\lambda}(\nu) + \mathbf{n}^i.
\end{equation}
In Eq.~\eqref{eqn:band_int_out_map_wMeff}: the sum over $\lambda$ spans different sky components, such as the CMB, dust, and synchrotron emission; the limits of integrations are the edges of the bandpass, assumed to be a top hat $\Delta\nu^i\equiv \nu^i_\text{max} - \nu^i_\text{min}$ and thus not made explicit; the input maps $\mathbf{s}^{\,i}_{in,\lambda}$ are smoothed with the beam\footnote{We consider achromatic circular Gaussian beams. Neglecting the beam frequency dependence ensures that the HWP and beam effects are fully decoupled. In contrast, other studies, such as Ref.~\cite{Duivenvoorden:2020xzm}, have modeled the coupling between non-ideal HWPs and beams.} of the corresponding frequency channel $i$; and $\mathbf{n}^i$ is the noise map contribution.
The effective Mueller matrix $\mathcal{M}_\text{eff}^i(\nu)$ can be approximated as
\begin{equation}\label{eqn:Meff}
    \mathcal{M}_\text{eff}^i(\nu) \simeq \begin{pmatrix}
        g^i(\nu) & 0 & 0 \\
        0 & \phantom{-}\rho^i(\nu) & \phantom{-}\eta^i(\nu) \\
        0 & -\eta^i(\nu) & \phantom{-}\rho^i(\nu) \\
    \end{pmatrix},
\end{equation}
where the non-zero elements are defined as\footnote{If the map-maker assumes the HWP to be ideal, the effective Mueller matrix elements reduce to the ones reported in Ref.~\cite{Monelli:2023wmv}, as expected.}:
\begin{subequations}\label{eqn:g_rho_eta}
\begin{align}
    \!\!\!\!g^i(\nu) & \equiv \frac
        {\text{\small{
            $2\widehat{m}_{II}m_{II}(\nu) + \widehat{m}_{QI}m_{QI}(\nu) + \widehat{m}_{UI}m_{UI}(\nu)$
        }}}
        {\text{\small{
            $2\widehat{m}_{II}^2 + \widehat{m}_{QI}^2 + \widehat{m}_{UI}^2$
        }}},\\[1mm]
    \!\!\!\!\rho^i(\nu) & \equiv \frac
        {\text{\small{
            $2\widehat{m}_{IQ}m_{IQ}(\nu) + \widehat{m}_{QQ}m_{QQ}(\nu) + \widehat{m}_{UQ}m_{UQ}(\nu) + 2\widehat{m}_{IU}m_{IU}(\nu) + \widehat{m}_{QU}m_{QU}(\nu) + \widehat{m}_{UU}m_{UU}(\nu)$
        }}}
        {\text{\small{
            $2\widehat{m}_{IQ}^2 + \widehat{m}_{QQ}^2 + \widehat{m}_{UQ}^2 + 2\widehat{m}_{IU}^2 + \widehat{m}_{QU}^2 + \widehat{m}_{UU}^2$
        }}},\!\!\!\\[1mm]
    \!\!\!\!\eta^i(\nu) & \equiv \frac
        {\text{\small{
            $2\widehat{m}_{IQ}m_{IU}(\nu) + \widehat{m}_{QQ}m_{QU}(\nu) + \widehat{m}_{UQ}m_{UU}(\nu) - 2\widehat{m}_{IU}m_{IQ}(\nu) - \widehat{m}_{QU}m_{QQ}(\nu) - \widehat{m}_{UU}m_{UQ}(\nu)$
        }}}
        {\text{\small{
            $2\widehat{m}_{IQ}^2 + \widehat{m}_{QQ}^2 + \widehat{m}_{UQ}^2 + 2\widehat{m}_{IU}^2 + \widehat{m}_{QU}^2 + \widehat{m}_{UU}^2$
        }}}.
\end{align}
\end{subequations}
In the equations above, the terms $m_{{SS}'}$ (with ${S,S}'=I,Q,U$) represent the frequency-dependent elements of the true Mueller matrix $\mathcal{M}_{\textsc{hwp}}^{\text{true}}$, while the coefficients $\widehat{m}_{SS'}$ correspond to the channel-dependent, band-integrated components of the estimated Mueller matrix ${m}_{SS'\!,\,\text{est}}$, obtained by weighting with the CMB SED in intensity units, $a_\textsc{cmb}(\nu)$, across the top hat bandpass:
\begin{equation}\label{eqn:m_ss}
    \widehat{m}_{SS'} \equiv \int_{\nu^i_\text{min}}^{\nu^i_\text{max}} \frac{\text{d}\nu}{\Delta\nu^i} \, a_{\rm CMB}(\nu) \, {m}_{SS'\!,\,\text{est}}(\nu)\,.
\end{equation}
The full derivation of the effective Mueller matrix is provided in Appendix~\ref{sec: appendix effective mueller}.

The formalism outlined above enables the injection of HWP systematics across multiple realizations of the desired data set through a fully map-based implementation, eliminating the need to generate simulated TODs followed by a map-making step. The validation of this approach is provided in Appendix~\ref{sec: appendix validation of map-based}.

Considering Eqs.~\eqref{eqn:band_int_out_map_wMeff} and~\eqref{eqn:g_rho_eta}, we highlight two points:
\begin{itemize}
    \item When $\mathcal{M}_\text{true}(\nu) \neq \mathcal{M}_\text{est.}(\nu)$, the output map $\mathbf{m}^i_\text{out}$ is inevitably distorted. This occurs because the two matrices modulate the sky signals differently within the band.
    \item When $\mathcal{M}_\text{true}(\nu) = \mathcal{M}_\text{est.}(\nu)$, we still expect some distortions due to the frequency-dependent scaling of the sky components. Recalling Eqs.~\ref{eq:cartoon map2tod_tod2map} and~\ref{eqn:band_int_out_map_wMeff}, the output map is given by the sum of two integrated terms, one accounting for the CMB signal ($\mathbf{s}_\textsc{cmb}$) and another for the foregrounds:
    \begin{equation}
    \label{eq:mismatch_comps}
        \mathbf{s}^i_\text{out} \!\simeq\! \mathbf{s}_\textsc{cmb} \underbrace{\int_{\nu^i_\text{min}}^{\nu_\text{max}}
        \frac{\text{d}\nu}{\Delta\nu^i}\, \mathcal{M}_\text{eff}^i(\nu)\alpha_{\text{I}}(\nu)\,}_\textrm{=1} + \int_{\nu^i_\text{min}}^{\nu_\text{max}}
        \frac{\text{d}\nu}{\Delta\nu^i}\, \mathcal{M}_\text{eff}^i(\nu) \, \alpha_{\text{I}}(\nu)\mathbf{s}_\textsc{fg}(\nu) + \mathbf{n}^i,
    \end{equation}
    where $\alpha_{\text{I}}(\nu)$ is the conversion factor\footnote{CMB maps are typically expressed in units of linearized differential temperature, representing fluctuations relative to the mean CMB temperature, $T_\text{CMB}$. The emitted intensity, defined as $I(\nu) = B_\nu(T=T_\text{CMB})$, corresponds to the brightness of a blackbody at frequency $\nu$ and temperature $T_\text{CMB}$. To express deviations from the mean temperature, $\Delta T_\text{CMB}$, we consider the linearization of the intensity:
    \begin{equation}
        dI(\nu) = \frac{dB_\nu(T_\text{CMB})}{dT_\text{CMB}} \, dT_\text{CMB} \quad \rightarrow \quad \Delta T_\text{CMB} = \frac{\Delta I(\nu)}{\left. \frac{dB_\nu(T)}{dT} \right|_{T = T_\text{CMB}}}=\frac{\Delta I(\nu)}{\alpha_{\text{I}}(\nu)}.
    \end{equation}
    Foreground maps are often provided in Rayleigh–Jeans units. However, in this analysis, all maps were converted and analyzed in CMB temperature units using the prescription above. If Rayleigh–Jeans units had been used, one could approximate the intensity in the Rayleigh–Jeans regime ($h\nu \ll k_{\rm B}T$) as $I(\nu) \simeq \frac{2k_{\rm B} \nu^2}{c^2} T_\text{CMB}$ \cite{Planck:2013wmz}. Therefore, the linearization approach outlined here remains applicable regardless of the original unit convention.} between brightness units and CMB ones. 
    While the CMB signal remains unperturbed due to its flat SED, the foreground signal, which varies with frequency, is still affected by the non-ideal response of the HWP.
\end{itemize}
Even with perfectly calibrated Mueller matrices, foreground emission with spatially varying SEDs can still be distorted by the instrumental response. While such distortions may, in principle, be mitigated if the true SEDs are known, removing them in practice would require either precise pixel-by-pixel SED knowledge or strong assumptions about foreground uniformity. Therefore, residual foreground distortions remain a potential source of systematic error even under ideal calibration assumptions.

\subsubsection{Sky models}
\label{sec:skys}
To simulate band-integrated maps according to Eq.~\eqref{eqn:band_int_out_map_wMeff}, we need sky maps as input. Since our model does not account for intensity-to-polarization leakage, we consider just the three main polarized sky components: the CMB and the Galactic synchrotron and thermal dust emissions. We simulate Gaussian realizations of the CMB signal using the best-fit $2018$ \textit{Planck} power spectra~\cite{Planck:2018vyg} ($TT$,$TE$,$EE$+low$E$+lensing). For the Galactic foreground sky, we instead adopt the Python Sky Model (\texttt{PySM}) package~\cite{Thorne:2016ifb}.
The dust~\cite{Planck:2015mvg} and synchrotron~\cite{Remazeilles:2014mba, Hinshaw_2013} sky signals are assumed to scale across frequencies, respectively, according to a modified black body (MBB) and power-law SED, which, in CMB units, are
\begin{subequations} \label{eqn:alambda}
\begin{align}
    a_{\mathrm{dust}}(\nu) &= 
        \left(\frac{\nu}{\nu_{\rm d}}\right)^{\beta_\mathrm{dust}-2}
        \frac{B_\nu(T_\mathrm{dust})}{B_{\nu_{\rm d}}(T_\mathrm{dust})}\frac{x_{\rm d}^2e^{x_{\rm d}}}{x^2e^{x}} 
        \frac{(e^{x}-1)^2}{(e^{x_{\rm d}}-1)^2}\,,\\
    a_{\mathrm{sync}}(\nu) &= 
        \left(\frac{\nu}{\nu_{\rm s}}\right)^{\beta_\mathrm{sync}-2} \frac{x_{\rm s}^2e^{x_{\rm s}}}{x^2e^{x}} 
        \frac{(e^{x}-1)^2}{(e^{x_{\rm s}}-1)^2}\,,
\end{align}
\end{subequations}
where $\nu_{\rm d}=337$ GHz and $\nu_{\rm s}=30$ GHz are the dust and synchrotron pivot frequencies, $B_\nu(T)$ denotes the black body spectrum at temperature $T$, $x_{(\rm d,s)}\equiv h\nu_{(\rm d,s)}/(k_{\rm B} T_0)$ and $T_0=2.725$ K is the average temperature of the CMB~\cite{Fixsen:2009ug}. We note that in the same units, the CMB has a flat SED, i.e., $a_\textsc{cmb}(\nu)=1$. In this work, we consider three different \texttt{PySM} models, namely \texttt{d0s0}, \texttt{d1s1} and \texttt{d10s5}, featuring increasing spatial variability of the foreground spectral parameters ($\beta_\mathrm{dust},\ T_\mathrm{dust}$, and $\beta_\mathrm{sync}$) across the sky and therefore increasing complexity.

\subsubsection{Noise}
A continuously rotating HWP modulates linear polarization to harmonics of the HWP rotation frequency, strongly suppressing low-frequency $1/f$ contamination in the reconstructed polarization maps. Modeling the map-level noise as white is therefore a good approximation for the present use-case~\cite{10.1063/5.0178066, Hill:2016jhd, Rashid:2023xnw}.
In particular, for each band the $Q$ and $U$ noise maps are simulated as independent Gaussian realizations with zero mean and variance set by the forecast sensitivity for that channel. This noise is injected after the optical/HWP chain, so that the propagation of HWP-induced systematics through the Mueller matrix is unaffected by the noise model~\cite{Monelli:2023wmv}.
This approximation holds under the following assumptions: (i) $1/f$ residuals are negligible in the polarization maps; (ii) the pixel-space noise covariance within a band is diagonal and spatially uniform; (iii) the noise is uncorrelated between detectors; and (iv) the scan is sufficiently cross-linked that hit-count inhomogeneities are negligible~\cite{Dupac:2001yi, Sutton:2008zh}.

More general cases can be accommodated without changing the analysis flow by replacing the diagonal map covariance with a pixel-by-pixel covariance to encode inhomogeneous depth or residual striping, and by introducing a detector-by-detector covariance to capture inter-detector correlations such as readout or common-mode atmospheric terms~\cite{Hamilton:2003xa}.
A potential coupling between HWP non-idealities and $1/f$ noise via detector/electronics nonlinearities can also imprint modulation-synchronous features and bias polarization if not mitigated. However, this effect is not modeled here and is deferred to future work.

\subsubsection{HWP specifics}
\label{subsec:hwp_syst}
We model HWP-related systematics following the procedure described in Ref.~\cite{Giardiello:2021uxq}, introducing mismatches between the true and estimated HWP parameters. For our simulations, the (estimated) HWP parameters are drawn from representative laboratory measurements~\cite{Komatsu_2021, Pisano_2022}, enabling us to quantify the impact of realistic calibration errors on scientific results. In particular, we consider optical systematics, which are described by the following HWP Jones matrix~\cite{ODea:2006tvb}:
\begin{equation} \label{eq: jones hwp}
    J_{\mathrm{HWP}}=\left(\begin{array}{cc}
    \left(1+h_1\right) e^{i \beta_1} & \zeta_1 e^{i \chi_1} \\
    \zeta_2 e^{i \chi_2}             & -\left(1+h_2\right) e^{i \beta_2}
\end{array}\right),
\end{equation}
where $h_{1,2}$ account for non-unitary-transmission, $\zeta_{1,2}$ and $\chi_{1,2}$ describe the magnitude and phase of cross-polarization, and $\beta_{1,2}$ capture deviations from the ideal $\pi$ phase-shift. All these parameters are real in general, and exactly null in the case of an ideal HWP.

In this analysis, we transition from the Jones to the Mueller formalism~\cite{Jones:2006ac, refId0} to more effectively capture systematic effects in CMB observations. Since the CMB signal is only partially polarized and CMB experiments record the incident intensity averaged over time, it is natural to describe the signal in terms of the Stokes vector $\mathbf{s} = (T, Q, U)$. Using the Mueller formalism facilitates the representation of these Stokes parameters and simplifies the tracking of systematic effects on each component of the polarization signal.

\paragraph{Injecting a miscalibration}
To account for the effect of miscalibrations in the HWP true parameters, we introduce Gaussian noise into the estimated frequency response of the non-ideal HWP. This is expressed as
\begin{equation} \label{eq: mismatch}
    \delta_\mathrm{true} = \delta_\mathrm{est} + \mathcal{N}(0, m \sigma),
\end{equation}
where $\delta$ represents one of the parameters of the HWP Jones matrix of Eq.~\eqref{eq: jones hwp} ($\delta = { h, \beta, \zeta, \chi }$), and $\mathcal{N}(0, m \sigma)$ denotes a sample from a Gaussian distribution with zero mean and standard deviation $m \sigma$. The parameter $\sigma$ represents the order of magnitude of the amplitude of the systematic effect, and $m = {0,1,2,3,4,5}$ is a multiplicative factor, called the \quotes{mismatch factor} throughout the text. The mismatch factor quantifies the amount of calibration uncertainty: $m=0$ indicates perfect calibration, while nonzero values of $m$ correspond to a specific uncertainty in the HWP characterization. Note that Eq.~\eqref{eq: mismatch} models stochastic measurement uncertainties as zero-mean Gaussian perturbations and does not account for systematic offsets, which may arise in realistic laboratory characterizations. Such offsets would introduce additional coherent bias terms that propagate through the entire analysis pipeline. While the present work focuses on stochastic mismatches, the impact of systematic biases is an important direction for future study.

Table~\ref{table: sigma mismatch for HWP syst} lists the specific values of $\sigma$ adopted in this work for each systematic effect. For cross-polarization, the value of $\sigma$ applies to both the real and imaginary parts of the systematic, rather than to $\zeta$ or $\chi$ individually.
\begin{table}
\begin{center}
\begin{tabular}{ l c } 
\hline
 \noalign{\smallskip} 
 HWP systematic $\delta$ & $\sigma$ $[\sqrt{\mathrm{GHz}}]$\\
 \noalign{\smallskip} 
 \hline
 \noalign{\smallskip} 
 Unit-transmission $h$ &  $1.90 \times 10^{-3}$ \\
 Phase-shift   $\beta$ &  $0.58^\circ$ \\
 Cross-polarization ($\Re[\zeta e^{i \chi}]$, $\Im[\zeta e^{i \chi}]$)&  $3.70 \times 10^{-4}$ \\
  \noalign{\smallskip} 
 \hline
\end{tabular}
\caption{Simulation of HWP systematic effects. The first column lists the HWP systematics' parameters defined in the Jones matrix (Eq.~\eqref{eq: jones hwp}), while the second column shows the corresponding order of magnitude for the miscalibration of each systematic (Eq.~\eqref{eq: mismatch}). The values reflect the representative calibration uncertainties adopted in the simulations, which were fine-tuned a posteriori to enable a meaningful exploration of their impact on $r$ inference. 
All entries carry units of $\sqrt{\rm{GHz}}$ because the HWP parameter perturbations are sampled with 1 GHz frequency resolution. Adopting a different sampling resolution would require rescaling $\sigma$ by the square root of the ratio of resolutions.}
\label{table: sigma mismatch for HWP syst}
\end{center}
\end{table}
\begin{figure}[ht!]
    \centering
    \includegraphics[width=.9\linewidth]{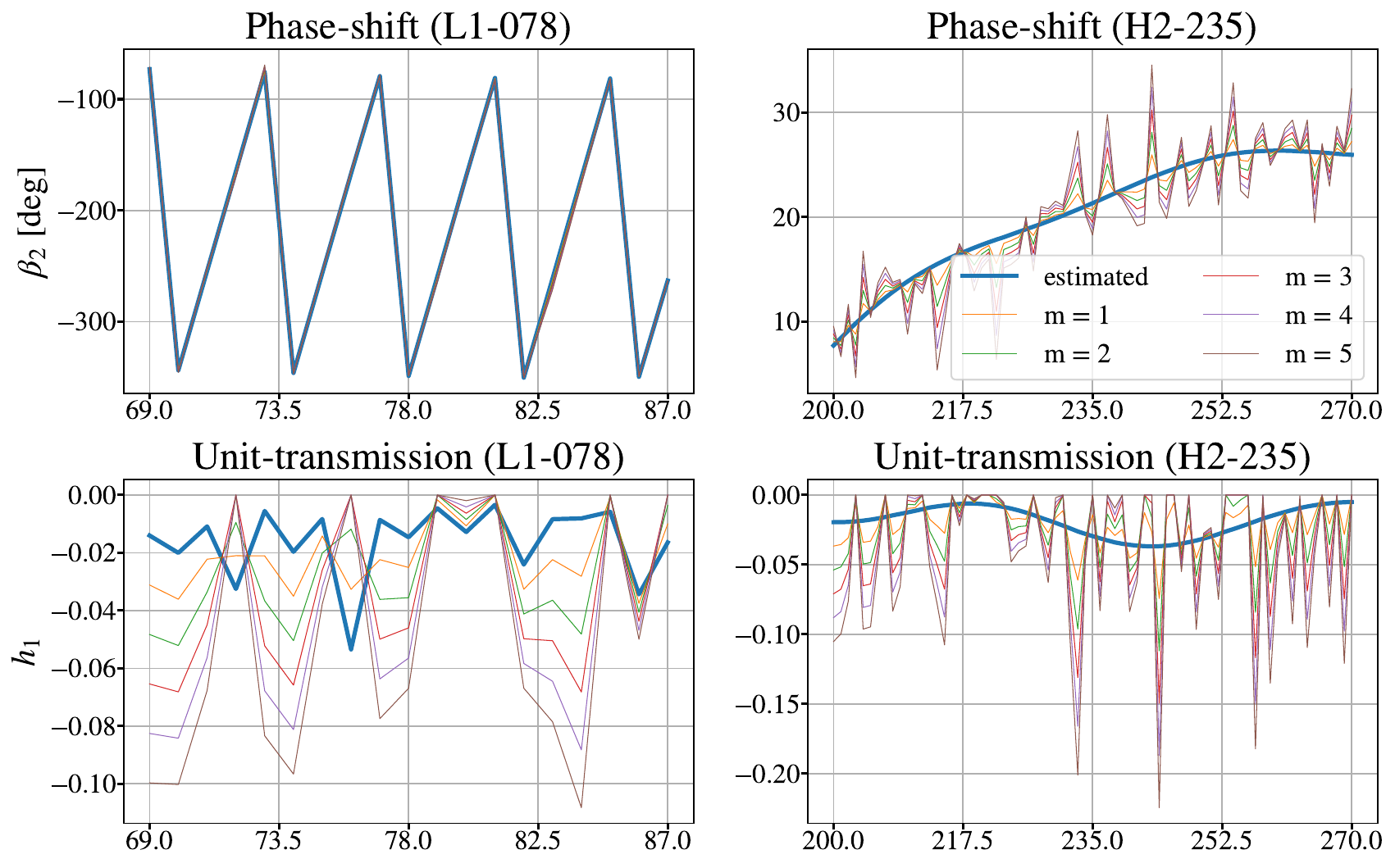}
    \includegraphics[width=.9\linewidth]{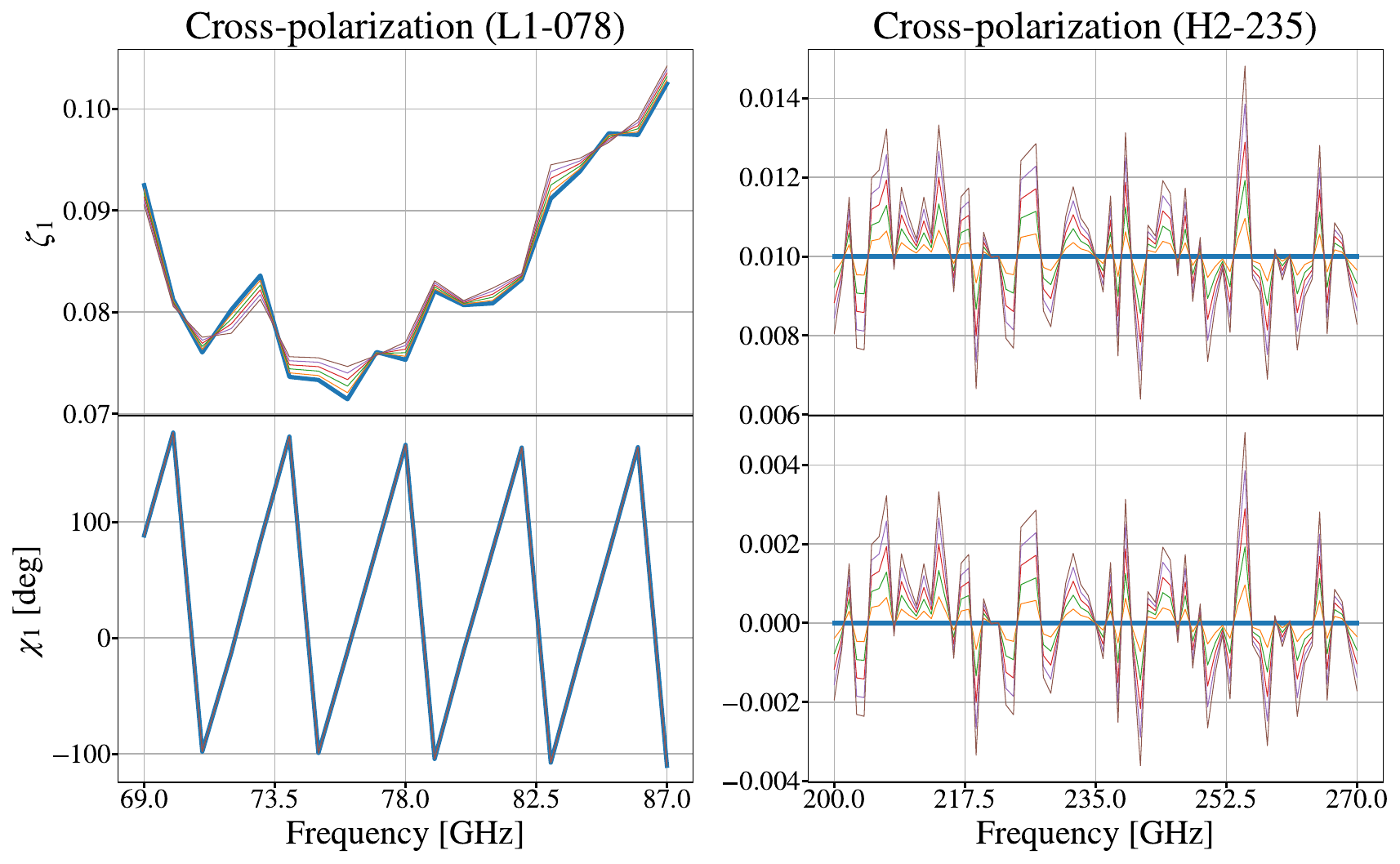}
    \vspace*{-5mm}
    \caption{Frequency profiles of HWP systematics for the two types of HWPs used by \textit{LiteBIRD}: the 5-layer achromatic HWP with anti-reflective coating for LFT (left)~\cite{Komatsu_2021}; and the metal-mesh HWP for HFT (right)~\cite{Pisano_2022}. The profiles are illustrated for the \quotes{L1-078} and \quotes{H2-235} channels. The rows display, in order: phase-shift (first), absorption (second), and cross-polarization module and phase (third and fourth). Each panel shows the estimated response (thick blue) alongside the true response for different miscalibrations ($m={1,2,3,4,5}$). Injection of the miscalibration is performed by fixing the seed (for the set of $m$ going from 1 to 5) per single systematic. Two distinct seeds are used for the real and imaginary parts of the cross-polarization.}
    \label{fig: mismatches for all syst}
\end{figure}
The mismatch between the estimated and true HWP responses is illustrated in figure~\ref{fig: mismatches for all syst}, which compares their frequency profiles. In this paper, we use two different types of HWP: a 5-layer achromatic HWP with an anti-reflective coating for the LFT (Low Frequency Telescope) instrument~\cite{Komatsu_2021} and a metal-mesh HWP for the HFT (High Frequency Telescope) instrument~\cite{Pisano_2022}. Figure~\ref{fig: mismatches for all syst} shows the HWP response for both an LFT channel (L1-078) and an HFT channel (H2-235), as an example, for each type of systematic. The differences between the two HWP models are evident. The HFT HWP (right panels) shows a smoother profile, with more visible deviations between true and miscalibrated cases. In contrast, the LFT HWP (left panels) exhibits a broken-line pattern with considerable variation, particularly in the phase parameters $\beta$ and $\chi$, and to a lesser extent in $h$ and $\zeta$. This pattern makes it difficult to visually distinguish between nearly overlapping curves.  
The strong and non-monotonic phase variation displayed in figure~\ref{fig: mismatches for all syst} distorts the reconstructed foreground signal, whose spectral energy distribution is integrated in frequency and modulated by the effective HWP Mueller matrix (see Eq.~\eqref{eqn:band_int_out_map_wMeff}). As we pointed out in section~\ref{subsec: sims}, even if we perfectly reconstruct and incorporate the knowledge of the HWP in the map-making step, the foreground signal would remain distorted. In the following, we will show that our likelihood isolates the incremental contamination from $m \ne 0$ relative to $m=0$, and therefore does not directly assess the absolute impact of the $m=0$ distortion on $r$; however, such frequency‑dependent distortions are expected to be largely absorbed by a minimum‑variance, blind component‑separation process that makes no assumptions about foreground SEDs or the HWP spectral shape.

We generate \cmbrealizations\ CMB and noise realizations, and for each realization, we perturb the non-ideal HWP parameters as described in Eq.~\eqref{eq: mismatch} considering the range of mismatch factors $m=1,2,3,4 \text{ and } 5$. For each realization, we use a fixed seed to introduce the distortion $\mathcal{N}(0, m \sigma)$; however, these are different for each systematic. We analyze two scenarios: a miscalibration affecting only one parameter in the Jones matrix of Eq.~\eqref{eq: jones hwp}, either unitary transmittance, cross-polarization, or phase shift; and a simultaneous miscalibration of all of them.


\subsection{Component separation} \label{subsec: comp sep}
\begin{figure}
    \centering
    \includegraphics[width=.8\linewidth]{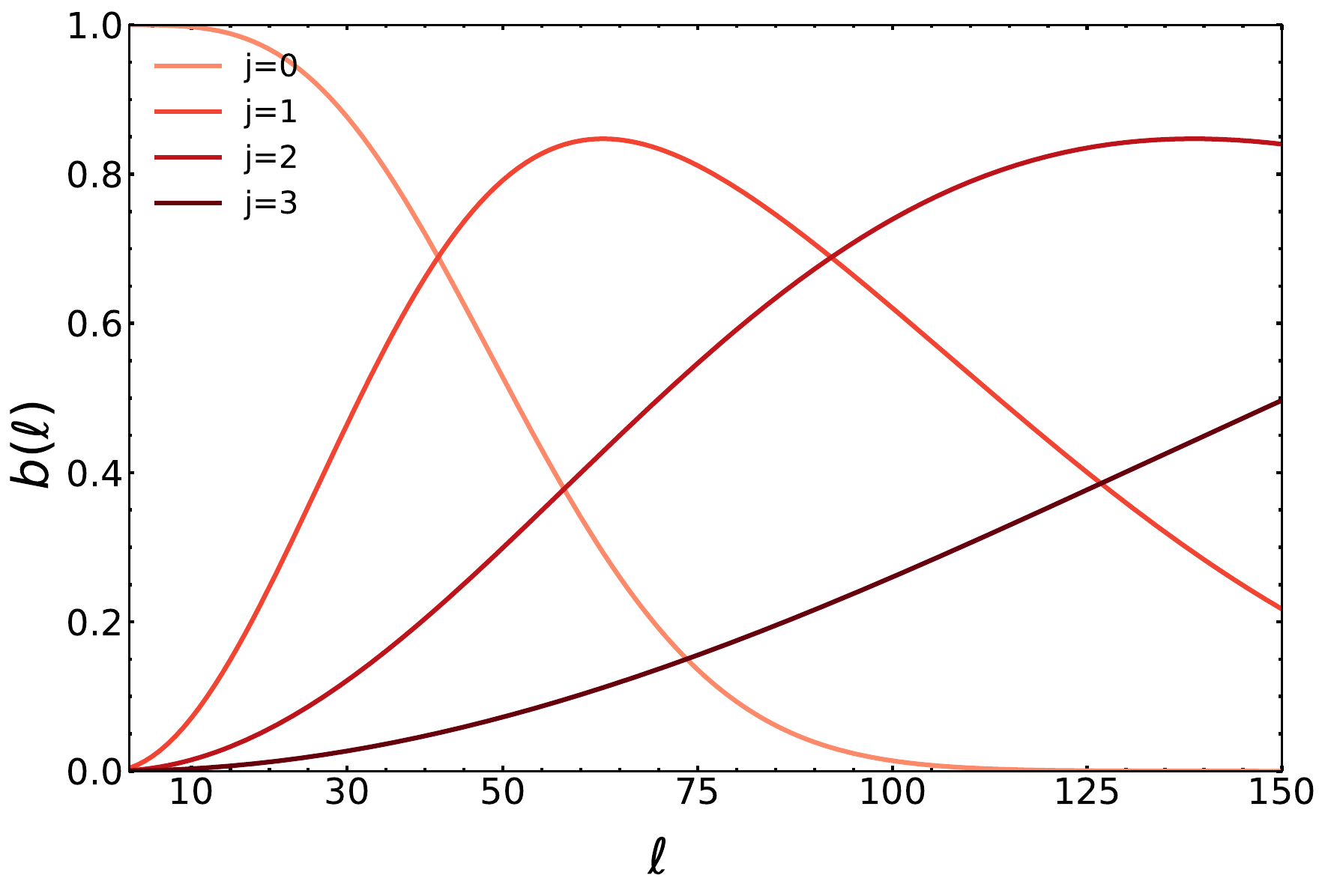}
    \caption{Configuration of needlet harmonic bands used for component separation in this work. Each color represents a different needlet scale, with bandpass filters $b(\ell)$ plotted as a function of multipole $\ell$.}
    \label{fig:needlets}
\end{figure}

To assess the impact of HWP miscalibration on the final estimate of the \ratio, we apply a component-separation algorithm to the simulated multi-frequency data set. In this work, we adopt the NILC methodology \cite{NILC_1}, which was widely used for the analysis of \textit{WMAP} and \textit{Planck} temperature data \cite{NILC_1,2013MNRAS.435...18B,Planck:2015mvg,Planck_nilc} and a promising avenue even for the processing of upcoming polarization data \cite{Carones:2022dnm,so_nilc,Carones:2022xzs}. NILC is a \quotes{blind} method, since it recovers the CMB signal without any modeling of the Galactic signal. This approach falls into the category of the minimum-variance pipelines, since it returns a foreground-cleaned CMB map by combining the data set channels with frequency-dependent weights so as to minimize the contribution of contaminants (such as foregrounds and noise) to the two-point statistics of the reconstructed cosmological field. To better mitigate these contaminants, the minimization is performed independently at different angular scales. In practice, this is achieved by filtering the harmonic coefficients of the multi-frequency data set using a set of harmonic bandpasses, known as needlets. For each bandpass, a set of multi-frequency needlet maps is then reconstructed. In this analysis, we use the so-called Mexican-hat needlet bands, shown in figure~\ref{fig:needlets}. 

As discussed in Sect.~\ref{sec:skys}, in the present work we consider different Galactic models with varying complexity. As shown in Ref.~\cite{Carones:2022xzs}, NILC effectively addresses the foreground contamination in a realistic simulated data set of the \textit{LiteBIRD} satellite, under the assumption of a fairly simple Galactic model (e.g., \texttt{d0s0}). However, for more complicated models, such as \texttt{d1s1} and \texttt{d10s5}, residual contamination may become significant, especially on the largest angular scales. Therefore, for these scales, all included in the first needlet band (shown in blue in figure~\ref{fig:needlets}), we employ an optimized NILC version, the MC-NILC~\cite{Carones:2022xzs}, to process \texttt{d1s1} and \texttt{d10s5} simulations. In MC-NILC, variance minimization is performed separately in different patches across the sky, with each patch corresponding to regions with similar spectral properties of the foregrounds. These properties are blindly traced by constructing the ratio of maps at two different \textit{LiteBIRD} frequency channels ($337$ and $119$ GHz). To avoid distortions in the sky partition caused by CMB and noise contamination, templates of foreground emission at the two frequencies are derived from the application of the Generalized NILC (GNILC) method~\cite{GNILC} on the simulated data set. This procedure for reconstructing the tracer is robust and model-independent, since the GNILC technique does not require any assumption on the underlying foreground emission, but only on the nuisance contribution of the CMB and instrumental noise to the estimated data covariance~\cite{GNILC}, which can generally be reliably predicted.  
In the context of $B$-mode analyses, the main source of uncertainty arises from the potential amplitude of the primordial tensor contribution in the CMB statistics; however, neglecting this contribution (i.e., assuming $r=0$ for the tracer reconstruction) is not expected to significantly affect the quality of the reconstructed tracer. Further details on the methodology and its robustness to different foreground models can be found in Ref.~\cite{Carones:2022xzs}.

Since the purpose of this work is to establish requirements on HWP miscalibration considering its effect on the estimate of $r$, we directly apply NILC, MC-NILC, and GNILC pipelines to $B$-mode maps constructed through a full-sky harmonic transformation of the simulated $Q$ and $U$ maps. All the adopted component-separation methodologies require input maps with a common angular resolution, while \textit{LiteBIRD} frequency channels have different beam sizes. Therefore, all maps, before being processed by component separation, are brought to a common resolution, which corresponds to that of the $40-$GHz largest beam (with $\textrm{FWHM} = 70.5$ arcmin).


\subsection{Power spectra estimation and likelihood maximization} \label{subsec: spectra and likelihood}
This section provides a description of the procedure we use to estimate the power spectra and the bias $\Delta r$ introduced by the HWP miscalibration.

Component separation is performed on both perfectly calibrated ($m=0$) and miscalibrated ($m \neq 0$) maps. The resulting maps include the CMB signal, residual foregrounds, and noise in the calibrated case ($\mathbf{s}_\mathrm{out}(m=0)$). In the miscalibrated one ($\mathbf{s}_\mathrm{out}(m \neq 0)$), an additional component is also present and originates from the propagation through component separation of the CMB distortions in the different frequency channels, introduced in Eq. \ref{eq:mismatch_comps} by the mismatch between $\mathcal{M}_\text{true}(\nu)$ and $\mathcal{M}_\text{est}(\nu)$. 

In both instances and for each simulation, we compute the $B$-mode angular power spectrum, using the \textit{Planck} \textit{GAL60} Galactic mask\footnote{\url{https://pla.esac.esa.int}} which retains 60\% of the sky, unless otherwise specified. The power spectra are estimated with pseudo-$C_\ell$s, properly corrected for power loss due to masking, and calculated using the \texttt{anafast}\footnote{\url{https://healpy.readthedocs.io/en/latest/generated/healpy.sphtfunc.anafast.html}} routine from the \texttt{healpy} public code~\cite{Gorski:2004by}. Although correlations between multipoles due to masking are not accounted for, this effect is considered negligible for large sky fractions, as discussed in Ref.~\cite{LiteBIRD:2022cnt}. 
Because we perform the $E/B$ decomposition on the full sky and apply masks only at the pseudo‑$C_\ell$ stage with the usual sky correction, the power‑spectrum estimation is not biased by any artificial $E$-to-$B$ leakage; HWP‑induced $Q/U$ mixing is instead captured by the model.
The $BB$ residual power spectra for both calibrated and miscalibrated cases, $C_{\ell}^{\text{res}}(m=0)$ and $C_{\ell}^{\text{res}}(m \neq 0)$, as modeled as
\begin{equation}\label{psres}
\begin{aligned} 
&C_{\ell}^\text{res}(m=0)      = C_{\ell}^\text{fg,res}(m=0)      + C_{\ell}^\text{n,res}(m=0),\\
&C_{\ell}^\text{res}(m \neq 0) = C_{\ell}^\text{fg,res}(m \neq 0) + C_{\ell}^\text{n,res}(m \neq 0) + C_{\ell}^\text{cmb,syst}(m \neq 0),
\end{aligned}
\end{equation}
where $C_{\ell}^\text{fg,res}$ and $C_{\ell}^\text{n,res}$ denote the power spectra of the residual foregrounds and noise after component separation, while $C_{\ell}^\text{cmb,syst}$ quantifies the contribution from systematic distortions affecting the reconstructed CMB signal, which arise only in the presence of miscalibration.

Following the work done in Ref.~\cite{LiteBIRD:2022cnt} and~\cite{Carralot2024_gain}, we adopt an exact likelihood~\cite{Gerbino:2019okg} in the harmonic domain to estimate the \ratio. This choice is justified by the large sky coverage adopted for power spectra computation in this work. The likelihood is expressed as
\begin{equation}
    -\ln \mathcal{L}\left(C_{\ell}{ }^{\mathrm{obs}} \mid r\right) = \sum_{\ell = \ell_\mathrm{min}}^{\ell_\mathrm{max}} \frac{2 \ell+1}{2} f_\text{sky}\left[\frac{C_{\ell}{ }^{\mathrm{obs}}}{C_{\ell}{ }^{\mathrm{th}}(r)}+\ln \left(C_{\ell}{ }^{\mathrm{th}}(r)\right)-\frac{2 \ell-1}{2 \ell+1} \ln \left(C_{\ell}{ }^{\mathrm{obs}}\right)\right],
\end{equation}
where we set the sky fraction $f_{\rm sky}=60\%$ ($50\%$ for \texttt{d1s1} and \texttt{d10s5}, see section~\ref{subsec:comb} for details), and $\ell_\mathrm{min}=2$, $\ell_\mathrm{max}=150$. We restrict our analysis to this narrower range of multipoles rather than using $\ell_\mathrm{max}=2\nside -1$, since our primary interest is in constraining the \ratio\ r. 
The observed and theoretical power spectra are defined as
\begin{equation}
\begin{split}
    &C_{\ell}^{\text{th}}(r)=r C_{\ell}^{\mathrm{GW}, r=1}+C_{\ell}^{\text{lensing}}+\langle C_{\ell}^\text{res}(m=0)\rangle,\\
    &C_{\ell}^{\text{obs}}(m= 0) = C_{\ell}^{\mathrm{res}}(m= 0) + C_{\ell}^{\mathrm{lensing}},\\
    &C_{\ell}^{\text{obs}}(m\neq 0) = C_{\ell}^{\mathrm{res}}(m\neq 0) + C_{\ell}^{\mathrm{lensing}},
\end{split}
\label{eq:theo_obs_cls}
\end{equation}
where $C_{\ell}^{\mathrm{GW},r=1}$ denotes the primordial gravitational wave $B$-mode power spectrum for $r=1$, $C_{\ell}^{\text{lensing}}$ is the lensing-induced $B$-mode power spectrum, and $C_{\ell}^{\text{obs}}(m)$ represents the observed power spectrum with ($m\neq0$) or without ($m=0$) HWP miscalibration. 
We include in the model $C_{\ell}^{\text{th}}(r)$ the average (over all simulations) of the residual power spectrum for the perfectly calibrated case ($\langle C_{\ell}^\text{fg,res}(m=0)\rangle+\langle C_{\ell}^\text{n}(m=0)\rangle$), since the goal of the analysis is to quantify through the cosmological parameter $r$ any deviation from the ideal (i.e., without miscalibration) power spectrum.   
Maximizing the likelihood, we estimate the best-fit value of $r$ for each simulation. The distortion of the \ratio\ $r$ due to the systematic effect is computed as the difference between the best-fit value in the miscalibrated and the ideal case:
\begin{equation}
    dr_m = r(m\neq 0) - r(m = 0),
    \label{eq:dr_m}
\end{equation}
where $m$ refers to the applied mismatch factor. The computation of such a difference, simulation by simulation, allows us to properly propagate only the impact of the systematic effect erasing the cosmic variance due to the specific considered realization. As done in Ref.~\cite{Carralot2024_gain}, the overall impact of the miscalibration is quantified with the rms, $\Delta r$, defined as:
\begin{equation}
    \Delta r (m) = \sqrt{\mu_{dr_m}^2 + \text{Var}[dr_m]} = \sqrt{\langle dr_m^2 \rangle},
    \label{eq:Delta_r}
\end{equation}
where $\mu_{dr_m}$ and $\text{Var}[dr_m]$ are the mean and variance of $dr_m$, to account for both the additional bias and variance on the \ratio. In Eq.~\eqref{eq:Delta_r} the angular brackets indicate averaging over the ensemble of simulations.

\subsection{Derivation of the requirements} \label{subsec: reqs}
\begin{figure}
    \centering\small
    \includegraphics[width=.5\linewidth]{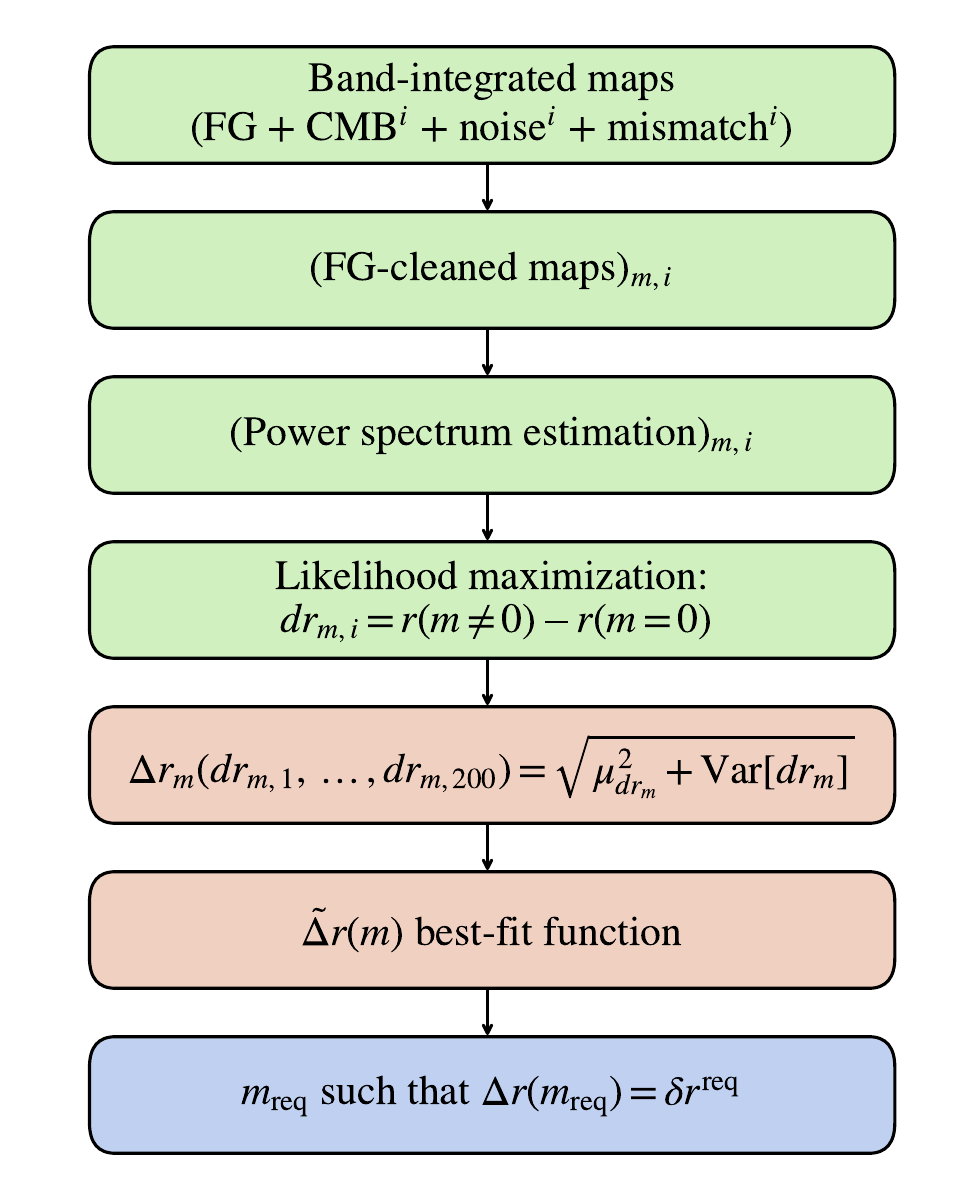}
    \vspace{-1mm}
    \caption{Methodology used in this study to determine the requirements for the mismatch factor, $m_\text{req}$, associated with HWP systematics, including non-unitary transmission, phase-shift, and cross-polarization, for a given foreground model.
    For each value of the mismatch factor $m \in \{1, 2, \dots, 5\}$, we simulate \cmbrealizations\ realizations (denoted by the index $i$ running over all realizations, represented by light green boxes) of the band-integrated maps and evaluate the distortion to the \ratio, $dr_{m,i}$, for each realization. These $dr_{m,i}$ values are then used to calculate the set of $\Delta r_m$ for different $m$ (shown in shaded pink), which is subsequently used to derive the best-fit function $\tilde{\Delta}r(m)$. The final step involves determining the value $m_\text{req}$ (light blue box), such that $\Delta r(m_\text{req}) = \delta r^\text{req} = 6.5 \times 10^{-6}$.}
    \label{fig:diagram}
\end{figure}

LiteBIRD design studies~\cite{LiteBIRD:2022cnt} allocate a budget to the uncertainty on the \ratio\ due to systematic instrumental effects, $\delta r^{\textrm{syst}}= 6.5 \times 10^{-4}$, and a requirement for any individual systematic effect to contribute less than one percent of the total budget, translating to $\delta r^{\textrm{req}} = 6.5 \times 10^{-6}$.
The goal of this paper is to determine the value of $m$ for each HWP systematic that satisfies this requirement. The procedure we follow is illustrated in figure~\ref{fig:diagram} and is outlined below:
\begin{enumerate}
\item We simulate multi-frequency \textit{LiteBIRD} data set with \cmbrealizations\ different CMB and noise realizations for $m={0,1,2,3,4,5}$ for a single HWP systematic miscalibration, as described in section~\ref{subsec:hwp_syst}. At this stage, the assumed foreground model is the \texttt{d0s0}. For this simplest case (\texttt{d0s0}), where all foreground components feature spatially-independent spectral parameters, the impact of HWP systematics could, in principle, be directly propagated into the power spectra analytically (see, e.g., Ref.~\cite{Monelli:2022pru}). However, since our main analysis pipeline propagates HWP distortions through map-based component separation, the spectrum-based approach is not directly applicable, and in any case cannot be extended to more realistic foreground scenarios, where the spatial variability of the foreground emission is taken into account.
\item The NILC pipeline is applied to each data set.
\item Injecting the corresponding observed and theoretical $B$-mode power spectra in the likelihood (computed as in Eq.~\eqref{eq:theo_obs_cls}), we obtain $dr_m$ (see Eq.~\eqref{eq:dr_m}) for each simulation. From the ensemble of simulations for $m={1,2,\dots,5}$, we compute $\Delta r$ (see Eq.~\eqref{eq:Delta_r}).
\item Using the resulting data points, we fit an analytical linear law $\tilde{\Delta}r(m)$ for $\Delta r$ as a function of $m$. From this, we derive the value $m_{\text{req}}$ for each HWP systematic, such that $\tilde{\Delta}r(m_{\text{req}}) = \delta r^{\textrm{req}}$.
\end{enumerate}

After having derived the requirements for unitary-transmission, phase-shift and cross-polarization individually, we inject all the three miscalibration effects through Eq. \ref{eqn:band_int_out_map_wMeff} simultaneously into a simulated data set, denoted as $\mathbf{s}_\mathrm{out}^{\mathrm{comb}}$, using the obtained $m_{\text{req}}$ values. We then apply the NILC pipeline to $\mathbf{s}_\mathrm{out}^{\mathrm{comb}}$ simulations and verify whether the combined effect, $\Delta r^{\mathrm{comb}}$, complies with the overall threshold, $\delta r^{\textrm{comb}} = 3 \times 6.5 \times 10^{-6} = 1.95 \times 10^{-5}$, derived under the assumption of totally uncorrelated effects. This combined analysis is repeated for the three different foreground models introduced in section~\ref{sec:skys} (\texttt{d0s0}, \texttt{d1s1} and \texttt{d10s5}).

In this analysis, the systematic effects were propagated simultaneously across all frequency channels to mimic a realistic observational scenario and to limit the number of required runs. Injecting the distortions collectively, rather than one channel at a time, provides a conservative estimate, since the resulting global requirement could be relaxed for bands where the systematic effect is weaker or more effectively mitigated by component separation.
\section{Results and discussion} \label{sec: results}

\subsection{Requirements on HWP systematics}
In this section, we report the outcome of the propagation of each HWP systematic effect (described in section~\ref{subsec:hwp_syst}) separately through the component-separation pipeline and its final impact on the \ratio, according to the procedure detailed in section~\ref{subsec: reqs}. This allows us to set requirements for the accuracy of the calibration of the HWP parameters. As already mentioned in section~\ref{subsec: reqs}, these requirements are derived from realistic \textit{LiteBIRD} simulations that assume a simple foreground model with constant spectral parameters across the sky (the PySM \texttt{d0s0}). Such a choice has been made in order to disentangle the sky complexity from the raw instrumental systematic effects. Nevertheless, we verify in section~\ref{subsec:comb} that the requirements derived in this simplified framework hold even for more realistic and complex scenarios.

\begin{figure}[ht!]
    \centering
    \includegraphics[width=.495\linewidth]{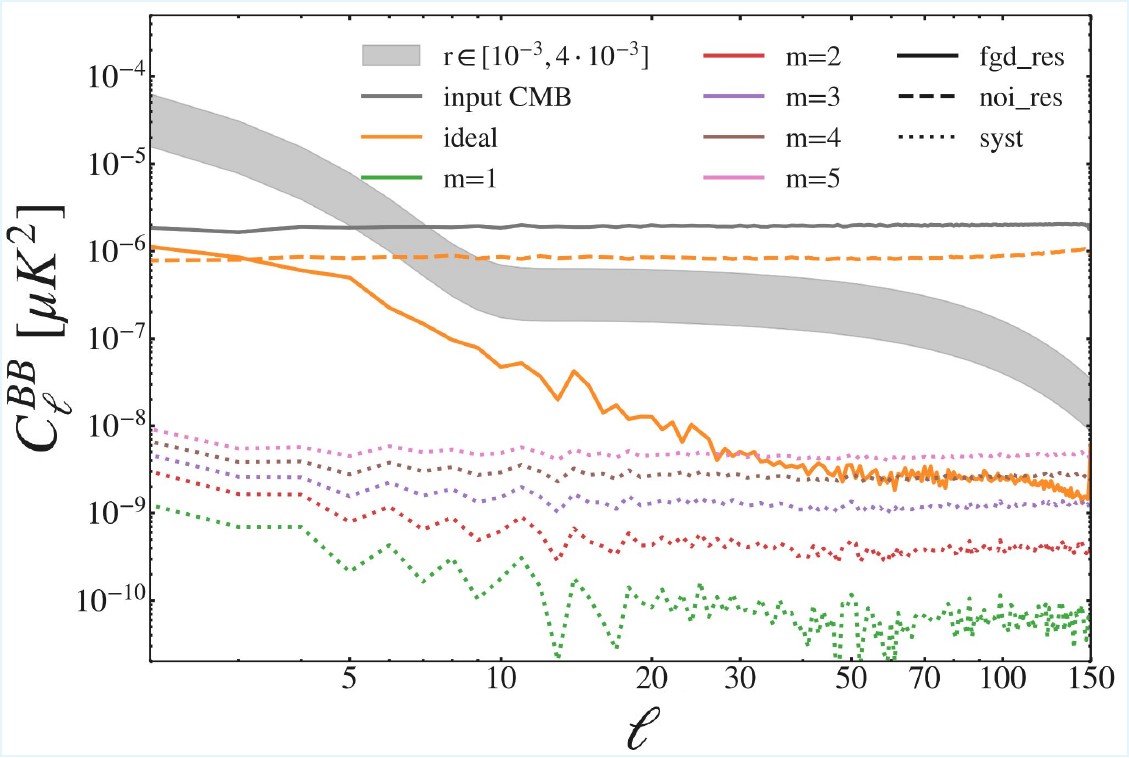}
    \raisebox{0.2cm}{\includegraphics[width=.495\linewidth]{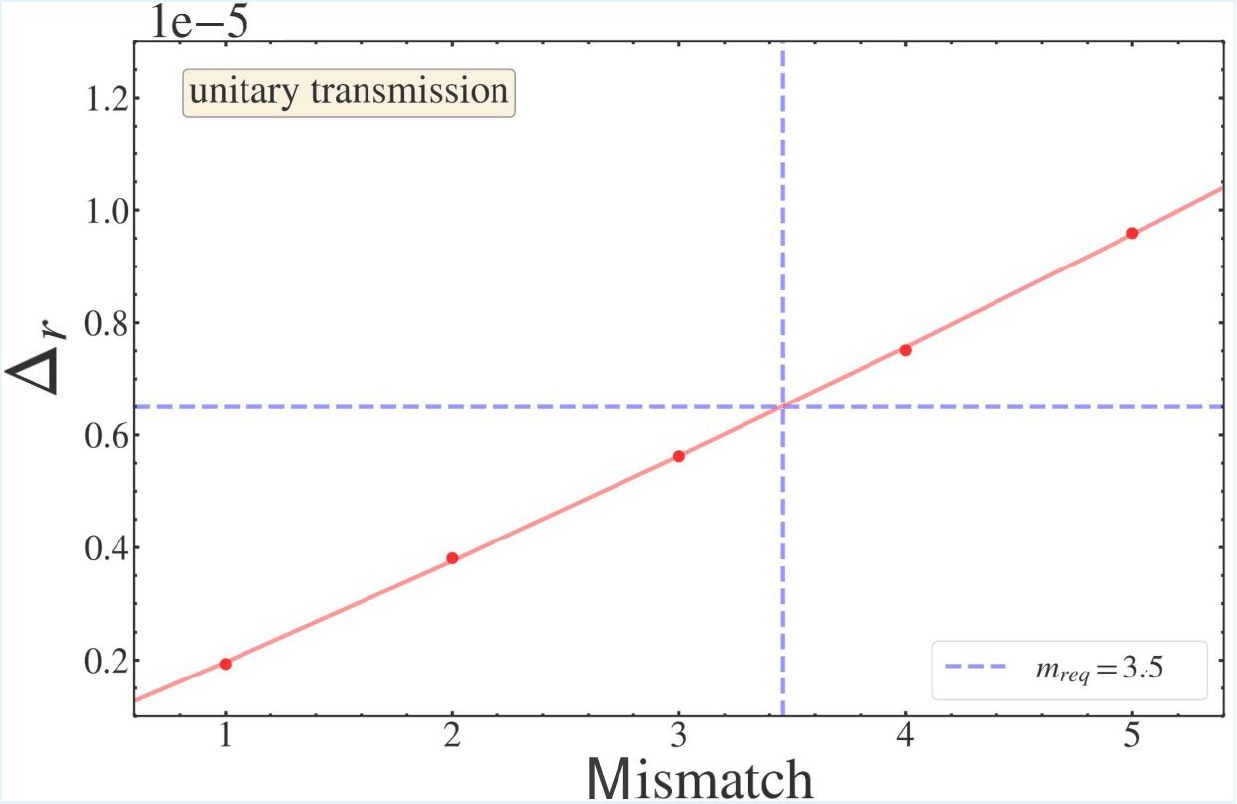}}
    \includegraphics[width=.495\linewidth]{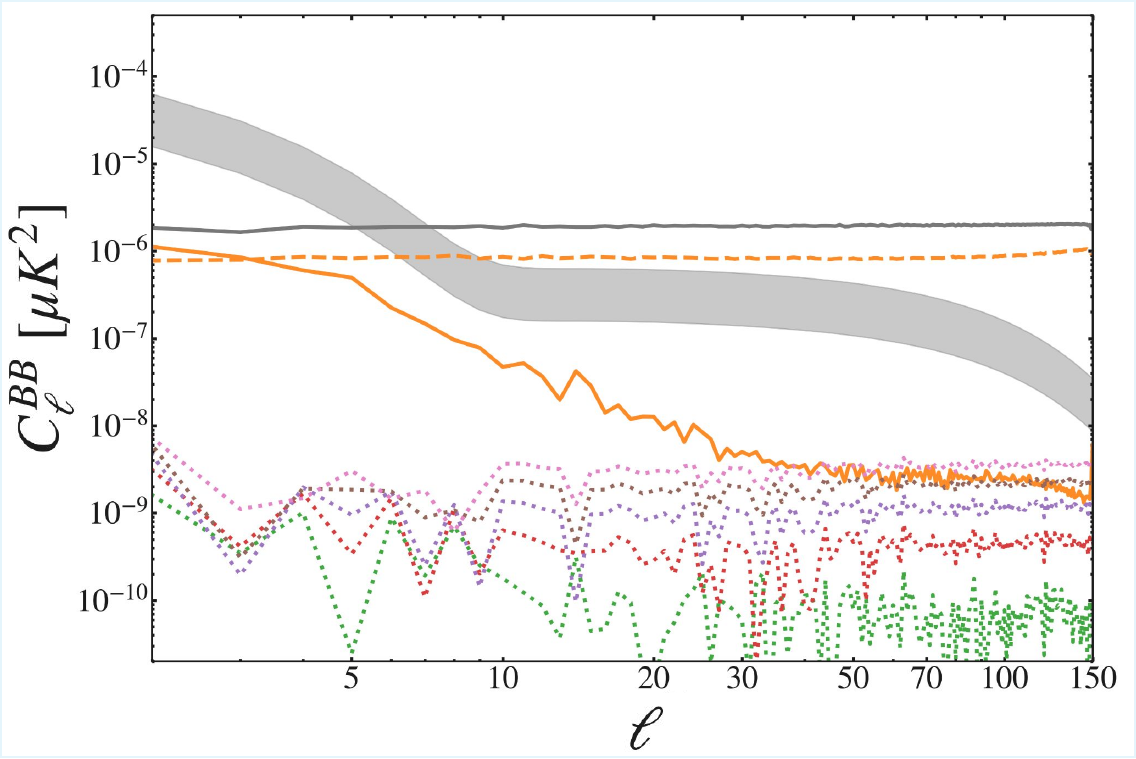}      
    \raisebox{0.2cm}{\includegraphics[width=.495\linewidth]{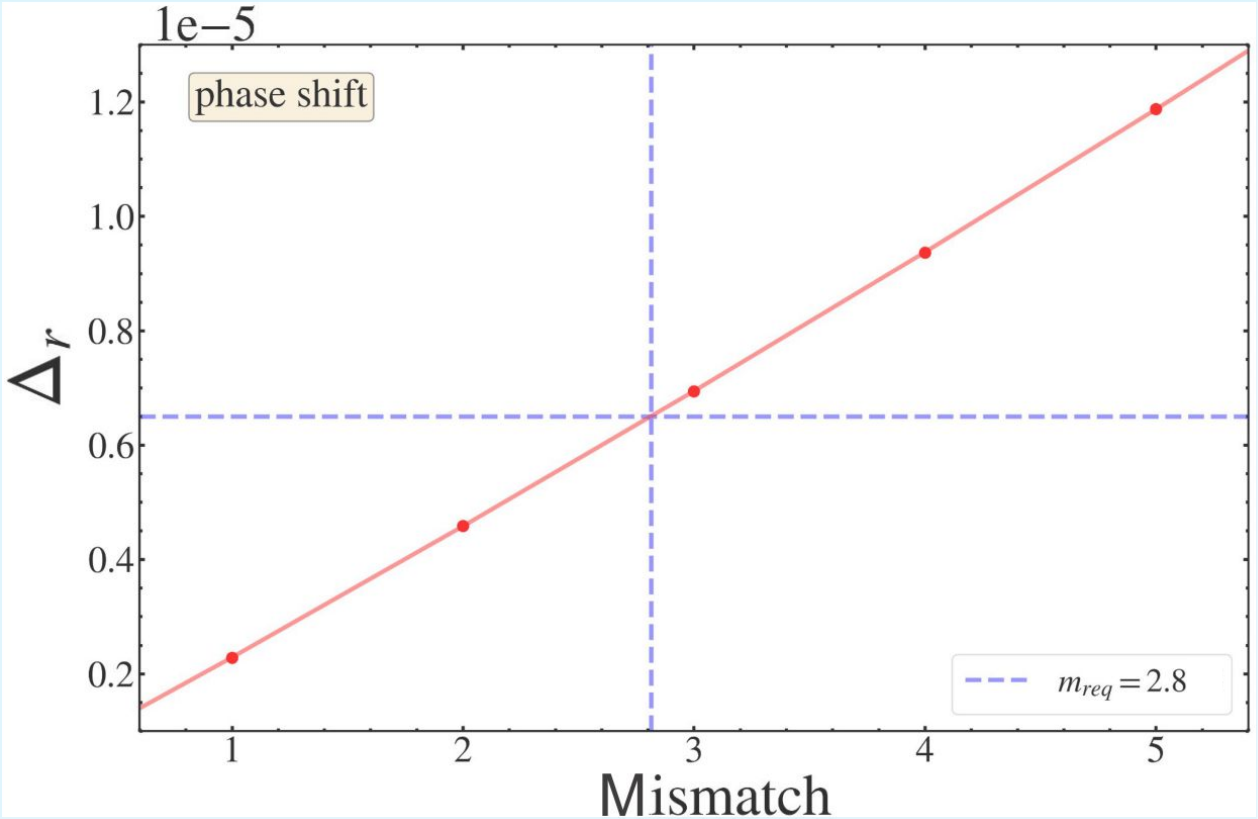}}
    \includegraphics[width=.495\linewidth]{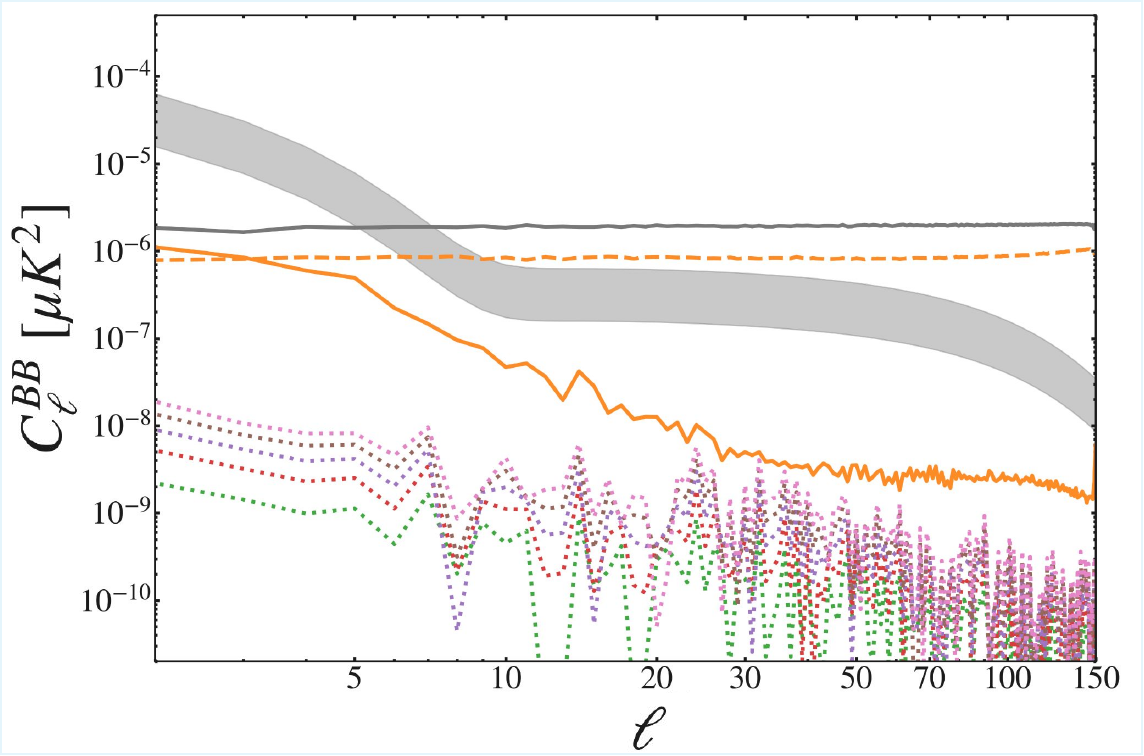}
    \raisebox{0.2cm}{\includegraphics[width=.495\linewidth]{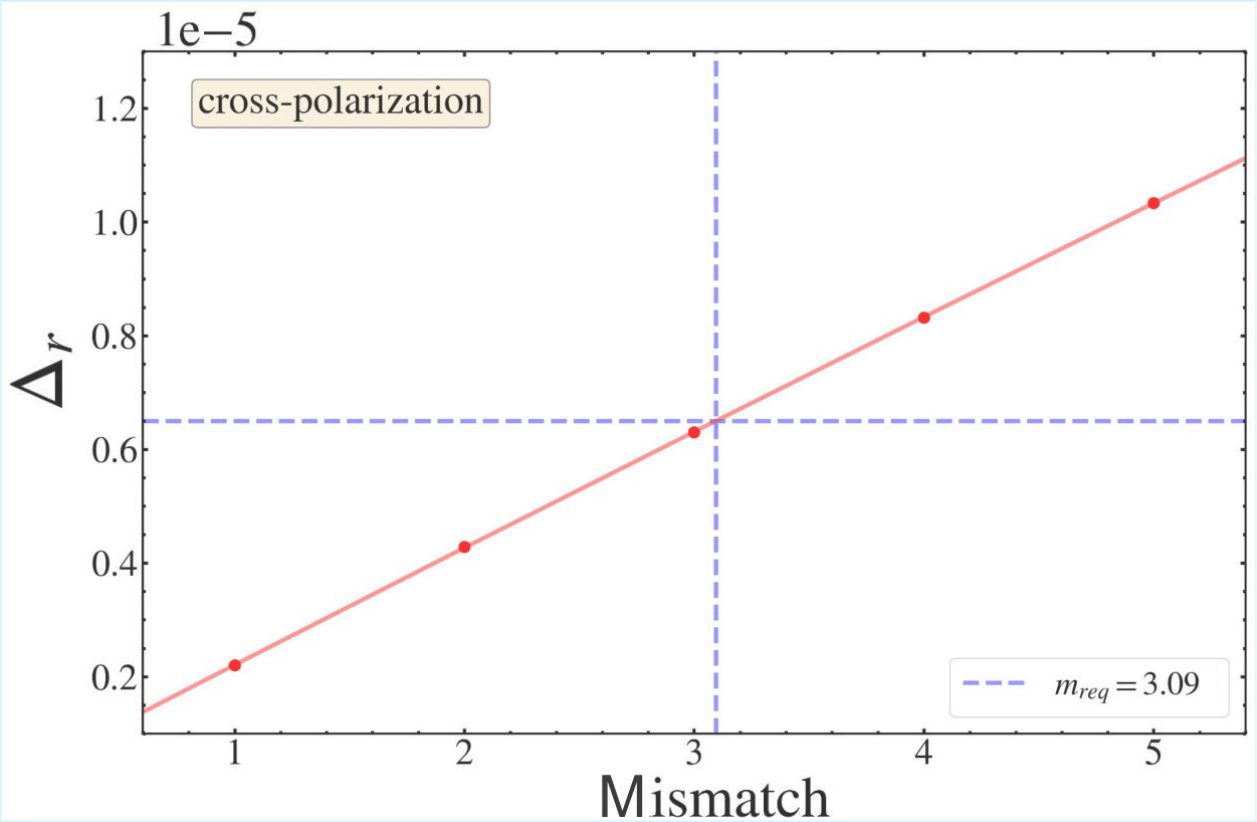}}
    \caption{Outcome of the propagation of the miscalibration of HWP unitary-transmission (top), phase-shift (middle) and cross-polarization (bottom) through the component-separation step. Left: average $B$-mode power spectrum (over \cmbrealizations\ simulations) of residual foregrounds and noise (solid and dashed orange lines) and systematic distortions (dotted) for different amplitude of the systematic effect ($m=0,1,...,5$). Foreground and noise residuals are reported just for the ideal case ($m=0$) because those affected by systematics would be superimposed when considering the adopted scales. The distortions' power spectrum is computed as the absolute value of the average difference between output angular power spectra with and without the systematic effect ($\left|\langle C_{\ell}^{\text{obs}}(m\neq0)-C_{\ell}^{\text{obs}}(m= 0)\rangle\right|$). For comparison, the mean input CMB $B$-mode power spectrum (grey solid line) and a range of primordial tensor $B$-mode spectra with $r\in [0.001,0.004]$ (gray shaded area) are shown. Right: trend of $\Delta r$ (computed according to Eq.~\eqref{eq:Delta_r}) as a function of the amplitude of the systematic mismatch factor $m$. The red solid lines represent the best-fit analytical expression, the blue dashed lines the threshold $\delta r^{\textrm{req}}=6.5\times 10^{-6}$ (horizontal), and the derived requirements $m_{\text{req}}$ (vertical).}
    \label{fig:cls_residuals}
\end{figure}

The NILC pipeline is applied to the different sets of simulations, each featuring a different kind of HWP miscalibration (cross-polarization, unitary-transmission, or phase-shift) and different amplitudes of the systematic effect (parametrized by the $m$ factor), as detailed in section~\ref{subsec:hwp_syst}. The average (over \cmbrealizations\ simulations) $B$-mode power spectrum of residual foregrounds, noise and overall instrumental systematic distortion (present in the cleaned CMB solution) are compared for the different $m$ values ($m=0,1,...,5$) and separately for each HWP systematic in the left panels of figure~\ref{fig:cls_residuals}. These different components contribute to the power spectrum of the total residuals in Eq.~\eqref{psres}. The systematic component is computed as $\left| C_{\ell}^{\text{obs}}(m\neq 0) - C_{\ell}^{\text{obs}}(m= 0)\right|$ and quantifies the overall distortion of the output power spectrum due to the systematic effect. We recall that, despite cases with different HWP systematics sharing the same set of $m$ values, the actual relative miscalibration uncertainties are not the same, as the corresponding $\sigma$ values in Eq.~\eqref{eq: mismatch} differ (see table~\ref{table: sigma mismatch for HWP syst}). 

The residual power spectrum for each simulation and for each $m$ and systematic case is injected in the likelihood as described in Sect.~\ref{subsec: spectra and likelihood} to derive the best-fit \ratio. We then collect the $m$-set of $dr_m$ (see Eq.~\eqref{eq:dr_m}) and compute $\Delta r(m)$ from Eq.~\eqref{eq:Delta_r}. The trend of $\Delta r$, as a function of $m$, for all types of HWP systematic miscalibration is shown in the right panels of figure~\ref{fig:cls_residuals}. 
We find that $\Delta r$ is well described by a linear function of $m$ for all the three systematic effects
We note that the residual power spectra in figure~\ref{fig:cls_residuals} (left panels) scale quadratically with the mismatch factor $m$, whereas the resulting bias $\Delta r$ (right panels) scales linearly. This distinction arises because the observed map is the sum of the sky signal and the systematic distortion, so the residual power spectrum contains both a linear cross-term (sky $\times$ systematic) and a quadratic auto-term (systematic $\times$ systematic). In the plotted spectra, which represent the ensemble average, the dominant linear cross-term averages to zero because the systematic is uncorrelated with the random CMB sky, leaving only the quadratic auto-term. In contrast, $\Delta r$ quantifies the rms (Eq.~\eqref{eq:Delta_r}), tracing the bias and the extra variance that is not shown in the plot. In this calculation, the variance is driven by the linear cross-term (which dominates over the quadratic term in any single realization due to the brightness of the sky signal) resulting in a linear scaling with $m$.

The analytical law $\tilde{\Delta}r(m)$ fitted to the observed $\Delta r$ values allows us to derive the requirements on $m$ (and therefore on the miscalibration uncertainties), $m_{\text{req}}$, for HWP cross-polarization, phase-shift and unitary-transmission. This is done by solving the equation $\tilde{\Delta}r(m_{\text{req}}) = \delta r^{\textrm{req}}$ with $\delta r^{\textrm{req}}=6.5\times 10^{-6}$. The obtained $m_{\text{req}}$ values are reported in the right panels of figure~\ref{fig:cls_residuals} and lead to the requirements on the HWP miscalibration uncertainties, $\sigma_{\text{req}}=m_{\text{req}}\sigma$, reported in table~\ref{table:HWP_requirements_d0s0}. Interestingly, the most stringent requirement refers to HWP cross-polarization, since it leaks the brighter $E$ modes into the observed $B$ modes. We note that the obtained results are similar to those found in Ref.~\cite{Giardiello:2021uxq}, despite the different analysis pipeline and the slightly different $\delta r^{\textrm{req}}$ considered in this work. Finally, we note that these requirements do not impose direct constraints on the HWP's physical construction, but rather on how well its non-idealities are characterized and incorporated into the analysis. This means that, as long as this approach is adopted, the HWP design can retain some flexibility, but precise laboratory-based characterization remains crucial.

\begin{table}
\begin{center}
\begin{tabular}{ l c c }
\hline
 \noalign{\smallskip} 
 HWP systematics & Requirement $\sigma_{\text{req}}$ (\texttt{d0s0}) $[\sqrt{\mathrm{GHz}}]$& Budget $\delta r^{\textrm{req}}$\\
  \noalign{\smallskip} 
 \hline
  \noalign{\smallskip} 
 Unit-transmission $h$ &  $\leq 6.56 \times 10^{-3}$ &\\
 Phase-shift   $\beta$ &  $\leq 1.63^\circ$ & $6.5\times 10^{-6}$\\
 Cross-polarization ($\Re[\zeta e^{i \chi}]$, $\Im[\zeta e^{i \chi}]$)&  $\leq 1.15 \times 10^{-3}$ \\
  \noalign{\smallskip} 
 \hline
\end{tabular}
\caption{Preliminary systematic effect requirements. The first column lists the HWP systematics defined in Eq.~\eqref{eq: jones hwp}, the second column shows the corresponding requirements on the calibration accuracy in order to meet the budget (third column), $\delta r^{\textrm{req}}$, for the single systematic miscalibration, assuming totally uncorrelated systematics for the simplest sky model. See the caption of Table~\ref{table: sigma mismatch for HWP syst} for an explanation of the $\sqrt{\mathrm{GHz}}$ dependance.}
\label{table:HWP_requirements_d0s0}
\end{center}
\end{table}

\subsection{Combination of HWP systematic effects}
\label{subsec:comb}
Once the accuracy requirements are individually determined for each HWP parameter, we assess the combined impact of all three HWP systematic effects on the estimation of the \ratio, $r$. To do this, we inject all three types of HWP miscalibrations into the same simulated data set, denoted as $\mathbf{s}_\mathrm{out}^{\mathrm{comb}}$, with mismatches corresponding to noise fluctuations (based on Eq.~\eqref{eq: mismatch}), with standard deviations equal to $\sigma_{\text{req}}$ as listed in table~\ref{table:HWP_requirements_d0s0}. As outlined in section~\ref{subsec: reqs}, in this combined scenario, the total budget on $\delta r$ to be fulfilled is $\delta r^{\textrm{comb}} = 3 \times 6.5 \times 10^{-6} = 1.95 \times 10^{-5}$.
\begin{figure}
    \centering
    \includegraphics[width=0.80\linewidth]{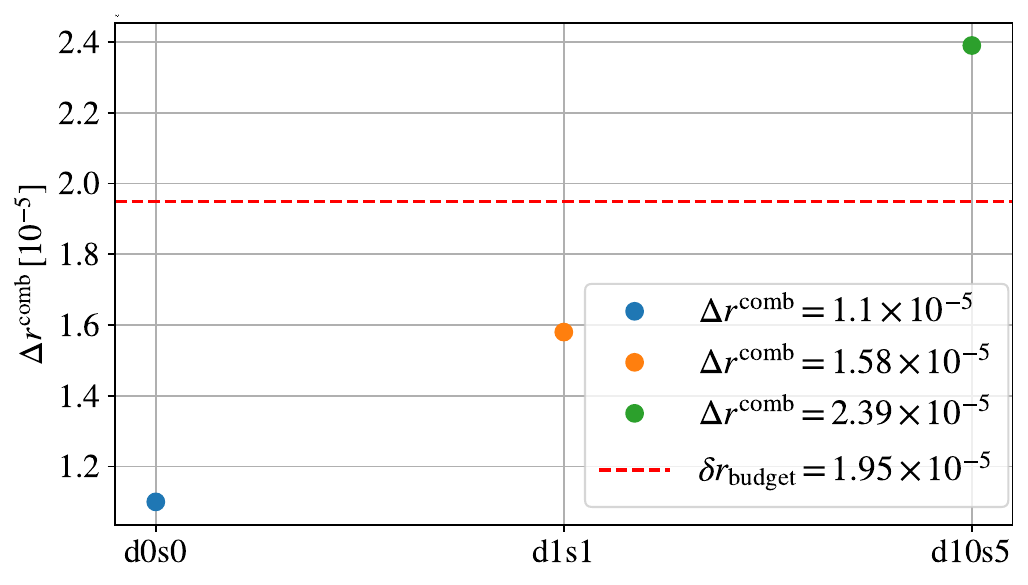}
    \caption{Trend of the bias on the tensor-to-scalar ratio $r$ as a function of the sky complexity, while simultaneously combining all the three kinds of systematic miscalibrations (see table~\ref{table:HWP_requirements_d0s0} for the miscalibration accuracy used).}
    \label{fig:deltar_comb}
\end{figure}
In figure~\ref{fig:deltar_comb}, we apply the NILC component-separation pipeline to \cmbrealizations\ different realizations of this combined data set, allowing us to estimate the combined systematic bias $\Delta r^{\text{comb}}$ due to the joint impact of all HWP miscalibrations. For the \texttt{d0s0} foreground model, we obtain $\Delta r^{\text{comb}}(\texttt{d0s0}) = 1.1 \times 10^{-5}$, as shown by the leftmost data point in figure~\ref{fig:deltar_comb}. This result covers $56\%$ of the total systematic budget of $\delta r_{\textrm{budget}}=1.95 \times 10^{-5}$, derived under the assumption of totally uncorrelated effects. This reduction occurs because the propagation of multiple systematic effects through the component-separation process is correlated. The NILC weights adjust to the combined distortions differently than they do when a single systematic miscalibration is considered. This phenomenon has been already observed in \textit{LiteBIRD} simulations, e.g., in the case of injection of multiple gain miscalibrations in different frequency channels, as reported in Ref.~\cite{Carralot2024_gain}.

This margin, relative to the total systematic budget $\delta r_{\textrm{budget}}$, allows us to assess the robustness of the calibration requirements in table~\ref{table:HWP_requirements_d0s0} against more complex sky models. This includes understanding how systematic effects interact with increasing foreground complexity. As noted in section~\ref{subsec: sims}, we generated combined data sets—each including the three HWP miscalibrations—based on the more challenging PySM \texttt{d1s1} and \texttt{d10s5} foreground models. Due to the increased complexity of these models, the component-separation pipeline and the masking strategy are accordingly adjusted to realistically handle such a challenging foreground subtraction. Specifically:
\begin{enumerate}
    \item in the first needlet band (shown in figure~\ref{fig:needlets}), we use MC-NILC instead of standard NILC, as described in section~\ref{subsec: comp sep};
    \item in addition to the \textit{Planck} \textit{GAL60} Galactic mask used in the \texttt{d0s0} analysis, we exclude an additional $10\%$ of the sky, corresponding to regions with high foreground residuals, resulting in a final sky fraction of $50\%$ (similar to the analyses in Ref.~\cite{LiteBIRD:2022cnt, Carones:2022xzs, Carralot2024_gain});
\end{enumerate}

Although the adopted masking strategy is not fully data-driven, previous work~\cite{Carones:2022xzs} has shown that realistic methods produce very similar results regarding residual amplitudes. We apply this modified pipeline to \textit{LiteBIRD} simulated data, including the combination of all the HWP systematics, with either the \texttt{d1s1} or \texttt{d10s5} foreground model. The results for $\Delta r^{\text{comb}}$ are shown in figure~\ref{fig:deltar_comb}, compared with the \texttt{d0s0} case. For the \texttt{d1s1} model, we find $\Delta r^{\text{comb}}(\texttt{d1s1}) = 1.58 \times 10^{-5}$, covering the $81\%$ of the total systematic budget. However, for the more complex \texttt{d10s5} model, $\Delta r^{\text{comb}}(\texttt{d10s5}) = 2.39 \times 10^{-5}$, essentially exceeding the available margin.

\begin{figure}
    \centering
    \includegraphics[width=1.0\linewidth]{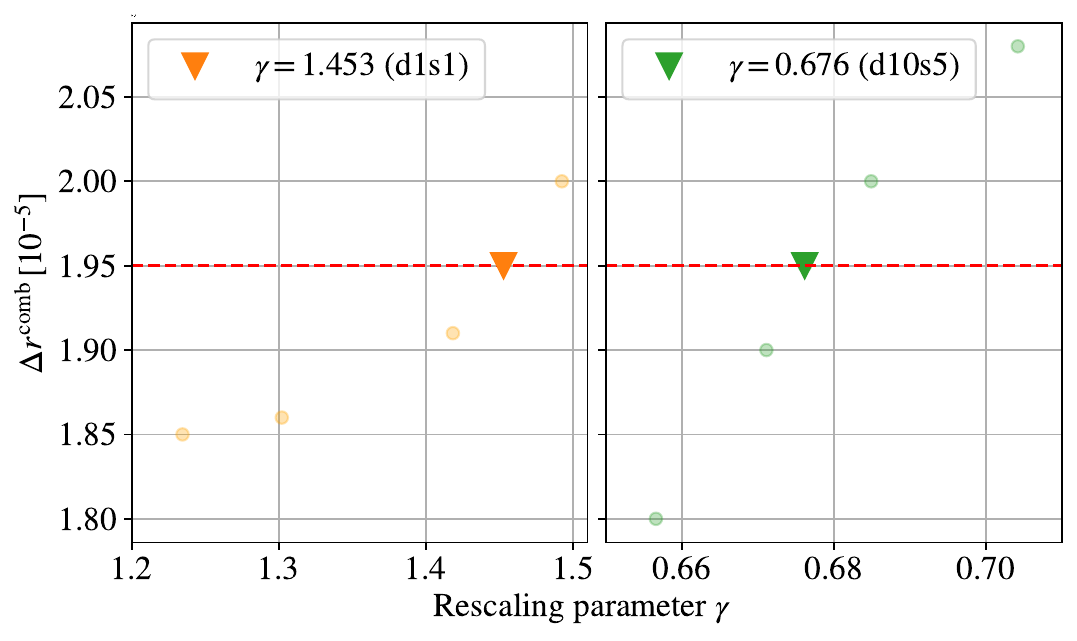}
    \caption{Trend of the bias on the tensor-to-scalar ratio $r$ (while simultaneously combining all the three kinds of systematic miscalibrations), for the most complicated sky models \texttt{d1s1} and \texttt{d10s5}, as a function of a rescaling parameter $\gamma$. Data points refer to Eq.~\eqref{eq:rescaling_factor}, and therefore to the progressive loosening/tightening of requirements on the HWP calibration accuracy for \texttt{d1s1}/\texttt{d10s5}. The orange dots refer to the \texttt{d1s1} case, while the green dots to the \texttt{d10s5} one and the red dashed line to the budget. The triangular dots provide the requirements to saturate the budget (see table~\ref{table:HWP_requirements_d1s1_d10s5}).}
    \label{fig:rescaling_factor}
\end{figure}

\paragraph{Rescaling requirements for \texttt{d1s1} and \texttt{d10s5} sky models}\quad
Since the requirements in table~\ref{table:HWP_requirements_d0s0} are quite loose for the \texttt{d1s1} sky model but not strict enough for \texttt{d10s5}, we aim to gradually rescale the miscalibration accuracy for each systematic effect in order to comply with the budget for these two scenarios. This is achieved by progressively relaxing or reducing the amplitude of each systematic miscalibration through a rescaling factor $\gamma$:
\begin{equation} \label{eq:rescaling_factor}
\Delta r^{\text{comb}} = \Delta r^{\text{comb}}(m_\mathrm{req} \, \sigma_\mathrm{syst} \, \gamma).
\end{equation}
The results of this scaling procedure for different values considered for the rescaling factor $\gamma$ are shown in figure~\ref{fig:rescaling_factor}. We find that, for \texttt{d1s1}, an increase of each systematic effect, corresponding to a scaling factor $\gamma = 1.453$, saturates the total budget $\Delta r=1.95 \cdot 10^{-5}$, while, for \texttt{d10s5}, the reduction corresponds to a factor of $\gamma = 0.676$. The left panel of figure~\ref{fig:rescaling_factor} shows a plateau in $\Delta r$ at low $\gamma$, consistent with a nonlinear regime where residuals and $\Delta r$ are affected by the coupling between HWP miscalibration and sky complexity, so further reductions in $\gamma$ yield diminishing returns. This is evident in the broader \texttt{d1s1} scan (left panel), whereas the restricted \texttt{d10s5} range (right panel) makes the effect harder to discern. A qualitatively similar saturation with calibration-like parameters is also observed in figure 7 of Ref.~\cite{Carralot2024_gain}.

We finally report the range of derived requirements for HWP systematic effects as a function of the foreground complexity in table~\ref{table:HWP_requirements_d1s1_d10s5}. In line with the discussion in Ref.~\cite{Carralot2024_gain}, we refer to the set of requirements obtained for the \texttt{d1s1} scenario as realistic, while that corresponding to \texttt{d10s5} framework as conservative.
\begin{table}
\begin{center}
\begin{tabular}{ lcc }
\hline
\noalign{\smallskip}
 HWP systematics requirement & $\sigma_{\text{req}}$ (\texttt{d1s1}) $[\sqrt{\mathrm{GHz}}]$& $\sigma_{\text{req}}$ (\texttt{d10s5}) $[\sqrt{\mathrm{GHz}}]$\\
 \noalign{\smallskip}
 \hline
 \noalign{\smallskip}
 Unit-transmission $h$ & $\leq 9.53 \times 10^{-3}$ & $\leq4.43 \times 10^{-3}$\\
 Phase-shift   $\beta$ & $\leq 2.37^\circ$ & $\leq1.10^\circ$ \\
 Cross-polarization ($\Re[\zeta e^{i \chi}]$, $\Im[\zeta e^{i \chi}]$) & $\leq 1.67 \times 10^{-3}$ & $\leq7.77 \times 10^{-4}$\\
 \noalign{\smallskip}
 \hline
\end{tabular}
\caption{Systematic effect requirements. The first column lists the HWP systematics defined in Eq.~\eqref{eq: jones hwp}, the second column shows the corresponding requirements on the calibration accuracy in order to meet the budget of $\delta r^{\textrm{req}} = 1.95 \times 10^{-5}$ up to a \texttt{d1s1} sky model. The third column displays instead the tightened requirements that comply with the budget for the most complicated sky model, \texttt{d10s5}. See the caption of Table~\ref{table: sigma mismatch for HWP syst} for an explanation of the $\sqrt{\mathrm{GHz}}$ dependance. }
\label{table:HWP_requirements_d1s1_d10s5}
\end{center}
\end{table}
\section{Conclusions} \label{sec: conclusions}
In this paper we have presented a comprehensive framework to assess and mitigate the impact of non-ideal half-wave plate (HWP) systematic effects on \textit{LiteBIRD}'s ability to measure the primordial $B$-mode polarization signal. Our approach introduces HWP non-idealities directly at the level of the observed maps and propagates their effects through an end-to-end pipeline that includes component separation (see Sect. \ref{subsec: comp sep}), power spectrum estimation, and likelihood-based inference of the \ratio, $r$ (see Sect. \ref{subsec: spectra and likelihood}).

We modeled the three primary HWP-induced systematics --- unitary transmission, phase-shift, and cross-polarization --- and introduced a parameterized miscalibration framework to evaluate their effects on $r$. For each individual systematic effect, we performed \cmbrealizations\ Monte Carlo simulations, varying the miscalibration amplitude and computing the induced bias $\Delta r$. The trend of $\Delta r$ as a function of the amplitude of each HWP systematic effect is reported in figure~\ref{fig:cls_residuals}. From these results, we derived calibration requirements that ensure each systematic effect, when treated in isolation and for the simplest sky model \texttt{d0s0}, contributes no more than $6.5 \times 10^{-6}$ to the bias on $r$, in accordance with \textit{LiteBIRD}'s allocated systematic budget. The obtained requirements are reported in table~\ref{table:HWP_requirements_d0s0}. The tightest constraint is found for cross-polarization, since it introduces direct $E$-to-$B$ leakage, and thus has the largest impact on $B$ modes.

We then studied the case in which all three HWP systematics are simultaneously miscalibrated at their respective threshold levels (see figure~\ref{fig:deltar_comb}). For the simplest foreground model, \texttt{d0s0}, the combined bias is found to be $\Delta r^{\text{comb}}(\texttt{d0s0}) = 1.1 \times 10^{-5}$, well below the total systematic budget of $1.95 \times 10^{-5}$ for the combination of three different miscalibration terms, thus proving that such systematic effects and their propagation and attenuation through component separation are correlated. Considering the available margin on the total $\Delta r$ in the \texttt{d0s0} scenario, we extended the analysis to two more complex Galactic models, \texttt{d1s1} and \texttt{d10s5}, in order to test the robustness of the derived requirements to the assumed foreground properties. In these more challenging scenarios, we incorporated improved foreground cleaning through the use of the MC-NILC component-separation technique~\cite{Carones:2022xzs} and additional sky masking to address residual contamination. As shown in figure~\ref{fig:deltar_comb}, for the \texttt{d1s1} case we find $\Delta r^{\text{comb}}(\texttt{d1s1}) = 1.58 \times 10^{-5}$, still within the allocated budget. However, for the most complex model, \texttt{d10s5}, the combined bias exceeded the budget, reaching $\Delta r^{\text{comb}}(\texttt{d10s5}) = 2.39 \times 10^{-5}$.

To address this, while still simultaneously miscalibrating all systematics, we introduced a scaling factor $\gamma$ (see Eq.~\eqref{eq:rescaling_factor} and figure~\ref{fig:rescaling_factor}) to progressively loosen/tighten the calibration accuracy for the \texttt{d1s1} and \texttt{d10s5} sky models. We demonstrated that the bias follows a non-linear trend in $\gamma$ when considering distinct sky models. We interpret this as a manifestation of the coupling between the HWP-induced distortions and the increasing foreground complexity. The recalibrated thresholds ensure that $\Delta r^{\text{comb}}$ remains within budget even in the presence of the most complex foregrounds. The final realistic and conservative calibration requirements, respectively for \texttt{d1s1} and \texttt{d10s5}, are summarized in table~\ref{table:HWP_requirements_d1s1_d10s5}.

Overall, our results demonstrate that \textit{LiteBIRD}'s target for controlling HWP systematics is achievable, provided that the non-idealities are accurately characterized and incorporated into the analysis.  Initial measurements of transmittance and phase using VNA (Vector Network Analyzer) setups show promising results, reaching the $10^{-2}$ level for transmission and approximately $1^\circ$ for phase, as demonstrated with Simons Observatory HWP measurements~\cite{Sugiyama_2024}. However, accurate cross-polarization measurements are still pending and the requirement of $10^{-3}$ presents a significant instrumental challenge.
Although we applied a uniform rescaling to all three systematic miscalibration parameters for simplicity, this choice was somewhat arbitrary. A more refined approach, where individual requirements are reassessed based on their relative impact on the science goal, may be more appropriate. For example, the stringent cross-polarization requirement might be moderately relaxed if compensated by a correspondingly tighter phase-shift constraint.
The derived requirements inform the required calibration precision during instrument testing and laboratory characterization, without imposing hard constraints on the HWP's physical construction.

However, while this study provides a comprehensive assessment of HWP-induced systematic effects and calibration requirements, some limitations should be noted. The analysis is conducted within the minimum-variance component-separation formalism, making no assumptions about foreground SEDs and the HWP spectral model. In practice, HWP parameters may be partially degenerate with foreground spectral parameters, complicating their disentanglement in a fully parametric framework. Simultaneous fitting of HWP and foreground properties—potentially within the mixing matrix—represents an avenue for further refinement, but is beyond the scope of this work.

Future works will extend this framework to include additional systematic effects, such as incidence-angle dependence and HWP-beam coupling, and to apply it in the context of full TOD-based end-to-end simulations. These efforts will further solidify \textit{LiteBIRD}'s capacity to reach its target sensitivity to primordial gravitational waves.

\acknowledgments
This work is supported in Japan by ISAS/JAXA for Pre-Phase A2 studies, by the acceleration program of JAXA research and development directorate, by the World Premier International Research Center Initiative (WPI) of MEXT, by the JSPS Core-to-Core Program of A. Advanced Research Networks, and by JSPS KAKENHI Grant Numbers JP15H05891, JP17H01115, and JP17H01125. The Canadian contribution is supported by the Canadian Space Agency. The French LiteBIRD phase A contribution is supported by the Centre National d’Etudes Spatiale (CNES), by the Centre National de la Recherche Scientifique (CNRS), and by the Commissariat á l’Energie Atomique (CEA). The German participation in LiteBIRD is supported in part by the Excellence Cluster ORIGINS, which is funded by the Deutsche Forschungsgemeinschaft (DFG, German Research Foundation) under Germany’s Excellence Strategy (Grant No. EXC-2094–390783311). The Italian LiteBIRD phase A contribution is supported by the Italian Space Agency (ASI Grants No. 2020-9-HH.0 and 2016-24-H.1-2018), the National Institute for Nuclear Physics (INFN) and the National Institute for Astrophysics (INAF). Norwegian participation in LiteBIRD is supported by the Research Council of Norway (Grant No. 263011) and has received funding from the European Research Council (ERC) under the Horizon 2020 Research and Innovation Programme (Grant agreement No. 772253 and 819478). The Spanish LiteBIRD phase A contribution is supported by the Spanish Agencia Estatal de Investigación (AEI), project refs. PID2019-110610RB-C21, PID2020-120514GB-I00, ProID2020010108 and ICTP20210008. Funds that support contributions from Sweden come from the Swedish National Space Agency (SNSA/Rymdstyrelsen) and the Swedish Research Council (Reg. no. 2019-03959). The US contribution is supported by NASA grant no. 80NSSC18K0132.
This work has also received funding by the European Union’s Horizon 2020 research and innovation program under grant agreement No 101007633 CMB-Inflate. Kavli IPMU is supported by World Premier International Research Center Initiative (WPI), MEXT, Japan. We acknowledge the use of computing facilities at CINECA.

\appendix
\section{Analytical Model: derivation}\label{sec: appendix effective mueller}
In this section we present the detailed derivation of Eqs.~\eqref{eqn:band_int_out_map_wMeff}, \eqref{eqn:Meff}, and \eqref{eqn:g_rho_eta}, which model the $(I,Q,U)_\text{out}$ maps reconstructed using a bin-averaging map-maker that assumes an HWP Mueller matrix different from the true one, under the assumption of perfect cross-linking, which means observing the same pixel from many distinct angles. For the sake of simplicity, we first assume pencil beams, single frequency, and no instrumental noise. We will then generalize to arbitrary Gaussian beams, finite frequency bands and include white noise.
\paragraph{Explicit derivation}
Assuming the instrument response to be linear, the noiseless TOD can be modeled as 
\begin{equation}\label{eqn:TODmodel}
    \mathbf{d}=A \; \mathbf{m}_\text{in}\,,
\end{equation}
where $\mathbf{m}_\text{in}$ represents the pixelized input sky maps and $A$ is the instrument response matrix. Given $\mathbf{d}$, the $(I,Q,U)_\text{out}$ maps obtained by bin-averaging are \cite{Tegmark:1996qs}
\begin{equation}\label{eqn:binning}
    \mathbf{m}_\text{out} = (\widehat{A}^\intercal \widehat{A})^{-1} \widehat{A}^\intercal  \mathbf{d}\,,
\end{equation}
where $\widehat{A}$ is the response matrix used by the map-maker.
By inserting the data model of Eq.~\eqref{eqn:TODmodel} into Eq.~\eqref{eqn:binning} and using that $A$ and $\widehat{A}$ are both sparse matrices with non-zero elements in the same entries, the binned (over time samples) Stokes vector $\mathbf{S}_{\text{out},p}$ at the pixel $p$ can be related to its input counterpart, $\mathbf{S}_{\text{in},p}$, by an effective Mueller matrix, $\mathcal{M}_{\text{eff},p}$, which is in principle pixel-dependent:
\begin{equation}\label{eqn:binningSp}
    \mathbf{S}_{\text{out},p} = \mathcal{M}_{\text{eff},p}\,\mathbf{S}_{\text{in},p} \equiv \left(\sum_{j't' \in \{jt\}_p} \widehat{\protect\fakebold{\mathbb{S}}}_{j't'} \widehat{\protect\fakebold{\mathbb{S}}}_{j't'}^\intercal \right)^{-1} \left(\sum_{jt \in \{jt\}_p} \widehat{\protect\fakebold{\mathbb{S}}}_{jt} \protect\fakebold{\mathbb{S}}_{jt}^\intercal  \right) \mathbf{S}_{\text{in},p}\,,
\end{equation}
where $\protect\fakebold{\mathbb{S}}_{jt}$ ($\widehat{\protect\fakebold{\mathbb{S}}}_{jt}$) encodes the true (estimated) instrument response of the detector $j$ at the time $t$ (and related to Eq.~\eqref{eq:optical_chain}). Eq.~\eqref{eqn:binningSp} can be equivalently written in terms of sky and instrument Stokes parameters as
\begin{align}\label{eqn:binningSS}
    \begin{pmatrix}
        {I} \\
        {Q} \\
        {U} \\
    \end{pmatrix}_{\text{out},p}
    &= \left[ \sum_{j't' \in \{jt\}_p} \begin{pmatrix}
        \widehat{\mathbbmsl{I}}^2 & 
        \widehat{\mathbbmsl{I}}\widehat{\mathbbmsl{Q}} &  
        \widehat{\mathbbmsl{I}}\widehat{\mathbbmsl{U}} \\
        \widehat{\mathbbmsl{Q}}\widehat{\mathbbmsl{I}} &
        \widehat{\mathbbmsl{Q}}^2 &   
        \widehat{\mathbbmsl{Q}}\widehat{\mathbbmsl{U}} \\
        \widehat{\mathbbmsl{U}}\widehat{\mathbbmsl{I}} & 
        \widehat{\mathbbmsl{U}}\widehat{\mathbbmsl{Q}} &  
        \widehat{\mathbbmsl{U}}^2 \\
        %
    \end{pmatrix}_{j't'} \right]^{-1}
    \left[ \sum_{jt\in\{jt\}_p}
    \begin{pmatrix}
        \widehat{\mathbbmsl{I}}\mathbbmsl{I} & 
        \widehat{\mathbbmsl{I}}\mathbbmsl{Q} &  
        \widehat{\mathbbmsl{I}}\mathbbmsl{U} \\
        \widehat{\mathbbmsl{Q}}\mathbbmsl{I} &
        \widehat{\mathbbmsl{Q}}\mathbbmsl{Q} &   
        \widehat{\mathbbmsl{Q}}\mathbbmsl{U} \\
        \widehat{\mathbbmsl{U}}\mathbbmsl{I} & 
        \widehat{\mathbbmsl{U}}\mathbbmsl{Q} &  
        \widehat{\mathbbmsl{U}}\mathbbmsl{U} \\
        %
    \end{pmatrix}_{jt} 
    \right]
    \begin{pmatrix}
        I \\
        Q \\
        U \\
    \end{pmatrix}_{\text{in},p}\nonumber\\
    &\equiv \begin{pmatrix}
        \braket{\widehat{\mathbbmsl{I}}^2} & 
        \braket{\widehat{\mathbbmsl{I}}\widehat{\mathbbmsl{Q}}} &  
        \braket{\widehat{\mathbbmsl{I}}\widehat{\mathbbmsl{U}}} \\
        \braket{\widehat{\mathbbmsl{Q}}\widehat{\mathbbmsl{I}}} &
        \braket{\widehat{\mathbbmsl{Q}}^2} &   
        \braket{\widehat{\mathbbmsl{Q}}\widehat{\mathbbmsl{U}}} \\
        \braket{\widehat{\mathbbmsl{U}}\widehat{\mathbbmsl{I}}} & 
        \braket{\widehat{\mathbbmsl{U}}\widehat{\mathbbmsl{Q}}} &  
        \braket{\widehat{\mathbbmsl{U}}^2} \\
        %
    \end{pmatrix}^{-1}
    \begin{pmatrix}
        \braket{\widehat{\mathbbmsl{I}}\mathbbmsl{I}} & 
        \braket{\widehat{\mathbbmsl{I}}\mathbbmsl{Q}} &  
        \braket{\widehat{\mathbbmsl{I}}\mathbbmsl{U}} \\
        \braket{\widehat{\mathbbmsl{Q}}\mathbbmsl{I}} &
        \braket{\widehat{\mathbbmsl{Q}}\mathbbmsl{Q}} &   
        \braket{\widehat{\mathbbmsl{Q}}\mathbbmsl{U}} \\
        \braket{\widehat{\mathbbmsl{U}}\mathbbmsl{I}} & 
        \braket{\widehat{\mathbbmsl{U}}\mathbbmsl{Q}} &  
        \braket{\widehat{\mathbbmsl{U}}\mathbbmsl{U}} \\
        %
    \end{pmatrix}
    \begin{pmatrix}
        I \\
        Q \\
        U \\
    \end{pmatrix}_{\text{in},p}\,,
\end{align}
where angled brackets denote averages over the set $\{jt\}_p$.

For an instrument with ideal detectors and whose first optical element is a rotating HWP with Mueller matix $\mathcal{M}_\textsc{hwp}$, the true instrument Stokes parameters are
\begin{subequations}\label{eqn:bbmsl}
    \begin{align}
        \mathbbmsl{I}_{j t} & =\frac{1}{2}\left[m_{II}+m_{QI} \mathrm{c}_{\beta_{j t}}+m_{UI} \mathrm{s}_{\beta_{j t}}\right], \\
        \mathbbmsl{Q}_{ j t} & =\frac{1}{2}\left[\left(m_{IQ}+m_{QQ} \mathrm{c}_{\beta_{ j t}}+m_{UQ} \mathrm{s}_{\beta_{ j t}}\right) \mathrm{c}_{\alpha_t}-\left(m_{IU}+m_{\textsc{qu}} \mathrm{c}_{\beta_{ j t}}+m_{UU} \mathrm{s}_{\beta_{j t}}\right) \mathrm{s}_{\alpha_t}\right], \\
        \mathbbmsl{U}_{j t} & =\frac{1}{2}\left[\left(m_{IQ}+m_{QQ} \mathrm{c}_{\beta_{j t}}+m_{UQ} \mathrm{s}_{\beta_{j t}}\right) \mathrm{s}_{\alpha_t}+\left(m_{IU}+m_{\textsc{qu}} \mathrm{c}_{\beta_{j t}}+m_{UU} \mathrm{s}_{\beta_{j t}}\right) \mathrm{c}_{\alpha_t}\right],
    \end{align}
\end{subequations}
where $\mathrm{s}_\theta \equiv \sin\theta$ and $\mathrm{c}_\theta \equiv \cos\theta$, $\alpha_t \equiv 2 \theta_t+2 \psi_t$ is twice the sum of the HWP ($\theta_t$) and telescope ($\psi_t$) angles, $\beta_{jt} \equiv 2 \theta_t-2 \xi_j$ is twice the difference of the detector ($\xi_i$) and HWP angles, and $m_{SS'}$ denote the elements of the HWP Mueller matrix. The estimated instrument Stokes parameters, $(\widehat{\mathbbmsl{I}},\widehat{\mathbbmsl{Q}},\widehat{\mathbbmsl{U}})_{jt}$, have the same structure as Eqs.~\eqref{eqn:bbmsl}, except that the HWP Mueller matrix elements contain the estimated $\widehat{m}_{SS'}$, rather than the true ones.

By inserting Eqs.~\eqref{eqn:bbmsl} and their estimated counterparts into Eq.\ \eqref{eqn:binningSS} and assuming perfect cross-linking (i.e., assuming the angles $\alpha_{jt}$ and $\beta_{jt}$ to be sampled uniformly enough in the set of observations $\{jt\}_p$ of the pixel $p$), some of the matrix elements appearing in Eq.~\eqref{eqn:binningSS} average to zero. The only non-vanishing terms are
\footnote{The rapid rotation of the HWP ensures that $\theta$ is uniformly sampled. Furthermore, for the approximation we just mentioned to work, $\xi+\psi$ should also be uniformly sampled, given the high number of detectors in \textit{LiteBIRD}, and the boresight rotation.}
\begin{subequations}\label{eqn:IIhathat}
\begin{align}
    \braket{\widehat{\mathbbmsl{I}}^2} &\simeq \frac{1}{4}\left[\widehat{m}_{II}^2 + \frac{\widehat{m}_{QI}^2 + \widehat{m}_{UI}^2}{2}\right]\,,\\
    \braket{\widehat{\mathbbmsl{Q}}^2} &\simeq \frac{1}{8}\left[\widehat{m}_{IQ}^2 + \frac{\widehat{m}_{QQ}^2 + \widehat{m}_{UQ}^2}{2} + \widehat{m}_{IU}^2 + \frac{\widehat{m}_{QU}^2 + \widehat{m}_{UU}^2}{2}\right]\,,\\
    \braket{\widehat{\mathbbmsl{U}}^2} &\simeq \frac{1}{8}\left[\widehat{m}_{IQ}^2 + \frac{\widehat{m}_{QQ}^2 + \widehat{m}_{UQ}^2}{2} + \widehat{m}_{IU}^2 + \frac{\widehat{m}_{QU}^2 + \widehat{m}_{UU}^2}{2}
    \right]\,,
\end{align}
\end{subequations}
and
\begin{subequations}\label{eqn:IIhat}
\begin{align}
    \braket{\widehat{\mathbbmsl{I}}\mathbbmsl{I}} &\simeq \frac{1}{4}\left[\widehat{m}_{II}m_{II} + \frac{\widehat{m}_{QI}m_{QI} + \widehat{m}_{UI}m_{UI}}{2}\right]\,,\\
    \braket{\widehat{\mathbbmsl{Q}}\mathbbmsl{Q}} &\simeq \frac{1}{8}\left[\widehat{m}_{IQ}m_{IQ} + \frac{\widehat{m}_{QQ}m_{QQ} + \widehat{m}_{UQ}m_{UQ}}{2} + \widehat{m}_{IU}m_{IU} + \frac{\widehat{m}_{QU}m_{QU} + \widehat{m}_{UU}m_{UU}}{2}\right]\,,\\
    \braket{\widehat{\mathbbmsl{Q}}\mathbbmsl{U}} &\simeq \frac{1}{8}\left[ \widehat{m}_{IQ}m_{IU} + \frac{\widehat{m}_{QQ}m_{QU} + \widehat{m}_{UQ}m_{UU}}{2} - \widehat{m}_{IU}m_{IQ} - \frac{\widehat{m}_{QU}m_{QQ} + \widehat{m}_{UU}m_{UQ}}{2}\right]\,,\\
    \braket{\widehat{\mathbbmsl{U}}\mathbbmsl{U}} &\simeq \frac{1}{8}\left[\widehat{m}_{IQ}m_{IQ} + \frac{\widehat{m}_{QQ}m_{QQ} + \widehat{m}_{UQ}m_{UQ}}{2} + \widehat{m}_{IU}m_{IU} + \frac{\widehat{m}_{QU}m_{QU} + \widehat{m}_{UU}m_{UU}}{2} \right]\,,\\
    \braket{\widehat{\mathbbmsl{U}}\mathbbmsl{Q}} &\simeq -\frac{1}{8}\left[\widehat{m}_{IQ}m_{IU} + \frac{\widehat{m}_{QQ}m_{QU} + \widehat{m}_{UQ}m_{UU}}{2} - \widehat{m}_{IU}m_{IQ} - \frac{\widehat{m}_{QU}m_{QQ} + \widehat{m}_{UU}m_{UQ}}{2}\right]\,.
\end{align}
\end{subequations}
One can then substitute these expressions into Eq.\ \eqref{eqn:binningSS} and write the effective Mueller matrix introduced in Eq.\ \eqref{eqn:binningSp} as\footnote{Note that, for a map-maker that assumes the HWP to be ideal, the effective Mueller matrix reduces to the one reported in~\cite{Monelli:2023wmv}, as expected.}
\begin{equation}\label{eqn:Meff_app}
    \mathcal{M}_\text{eff} \simeq \begin{pmatrix}
        g & 0 & 0 \\
        0 & \rho & \eta \\
        0 & -\eta & \rho \\
    \end{pmatrix},
\end{equation}
where we have defined
\begin{subequations}\label{eqn:grhoeta_app}
\begin{align}
    g & \equiv \frac
    {2\widehat{m}_{II}m_{II}  + \widehat{m}_{QI}m_{QI}  + \widehat{m}_{UI}m_{UI}}
    {2\widehat{m}_{II}^2 + \widehat{m}_{QI}^2 + \widehat{m}_{UI}^2},\\
    \rho & \equiv \frac
    {2\widehat{m}_{IQ}m_{IQ}  + \widehat{m}_{QQ}m_{QQ}  + \widehat{m}_{UQ}m_{UQ}  + 2\widehat{m}_{IU}m_{IU}  + \widehat{m}_{QU}m_{QU}  + \widehat{m}_{UU}m_{UU} }
    {2\widehat{m}_{IQ}^2 + \widehat{m}_{QQ}^2 + \widehat{m}_{UQ}^2 + 2\widehat{m}_{IU}^2 + \widehat{m}_{QU}^2 + \widehat{m}_{UU}^2},\\
    \eta & \equiv \frac
    {2\widehat{m}_{IQ}m_{IU}  + \widehat{m}_{QQ}m_{QU}  + \widehat{m}_{UQ}m_{UU}  - 2\widehat{m}_{IU}m_{IQ}  - \widehat{m}_{QU}m_{QQ}  - \widehat{m}_{UU}m_{UQ}}
    {2\widehat{m}_{IQ}^2 + \widehat{m}_{QQ}^2 + \widehat{m}_{UQ}^2 + 2\widehat{m}_{IU}^2 + \widehat{m}_{QU}^2 + \widehat{m}_{UU}^2}.
\end{align}
\end{subequations}
Going from Eqs.~\eqref{eqn:Meff_app} and \eqref{eqn:grhoeta_app} to Eqs.~\eqref{eqn:band_int_out_map_wMeff}, \eqref{eqn:Meff} and \eqref{eqn:g_rho_eta} is straightforward from these steps.
\begin{itemize}[parsep=0pt]
    \item Gaussian beams can be included in the model by smoothing the input maps by the appropriate full-width-half-maximum\footnote{Assuming circular and achromatic beams allows us to decouple beam convolution from the HWP response. Conversely, such an approximation is relaxed in Ref.~\cite{Duivenvoorden:2020xzm}.}.
    \item To go beyond the single frequency assumption, we allow the HWP response to depend on frequency, i.e.\ substitute $m_{SS'}$ with $m_{SS'}(\nu)$. Furthermore, we redefine the estimated $\widehat{m}_{SS'}$ as the average of $\widehat{m}_{SS'}(\nu)$ over the frequency band.
    \item White noise is added as an independent Gaussian contribution to the $\mathbf{m}_\text{out}$ maps after the optical and HWP systematic effects. More general noise models, including correlated or inhomogeneous noise, would likewise enter at this stage without qualitatively changing our conclusions on the approximation's validity. A detailed investigation of possible couplings between HWP systematics and realistic noise properties is beyond this work's scope and deferred to future studies.
\end{itemize}
With these changes, and writing explicitly the sum over different sky components, we recover Eqs.~\eqref{eqn:band_int_out_map_wMeff}, \eqref{eqn:Meff}, and \eqref{eqn:g_rho_eta}.


\section{Analytical Model: validation}\label{sec: appendix validation of map-based}
Because of the scanning strategy specifics and the fast rotation of the HWP, we expect \textit{LiteBIRD} to have sufficient cross-linking and Eqs.~\eqref{eqn:band_int_out_map_wMeff}, \eqref{eqn:Meff}, and \eqref{eqn:g_rho_eta} to hold. To make sure that this is the case, we test their predictions against TOD-based simulations. We carried on this test for all the values of the mismatch factor we consider, $m\in \{1,2,\dots,5\}$. Since the result we obtain do not depend on $m$, here we report the outcome of the validation for perfect calibration only, i.e., $m=0$.

\begin{figure}[!ht]
    \centering
    \includegraphics[width=1.\linewidth]{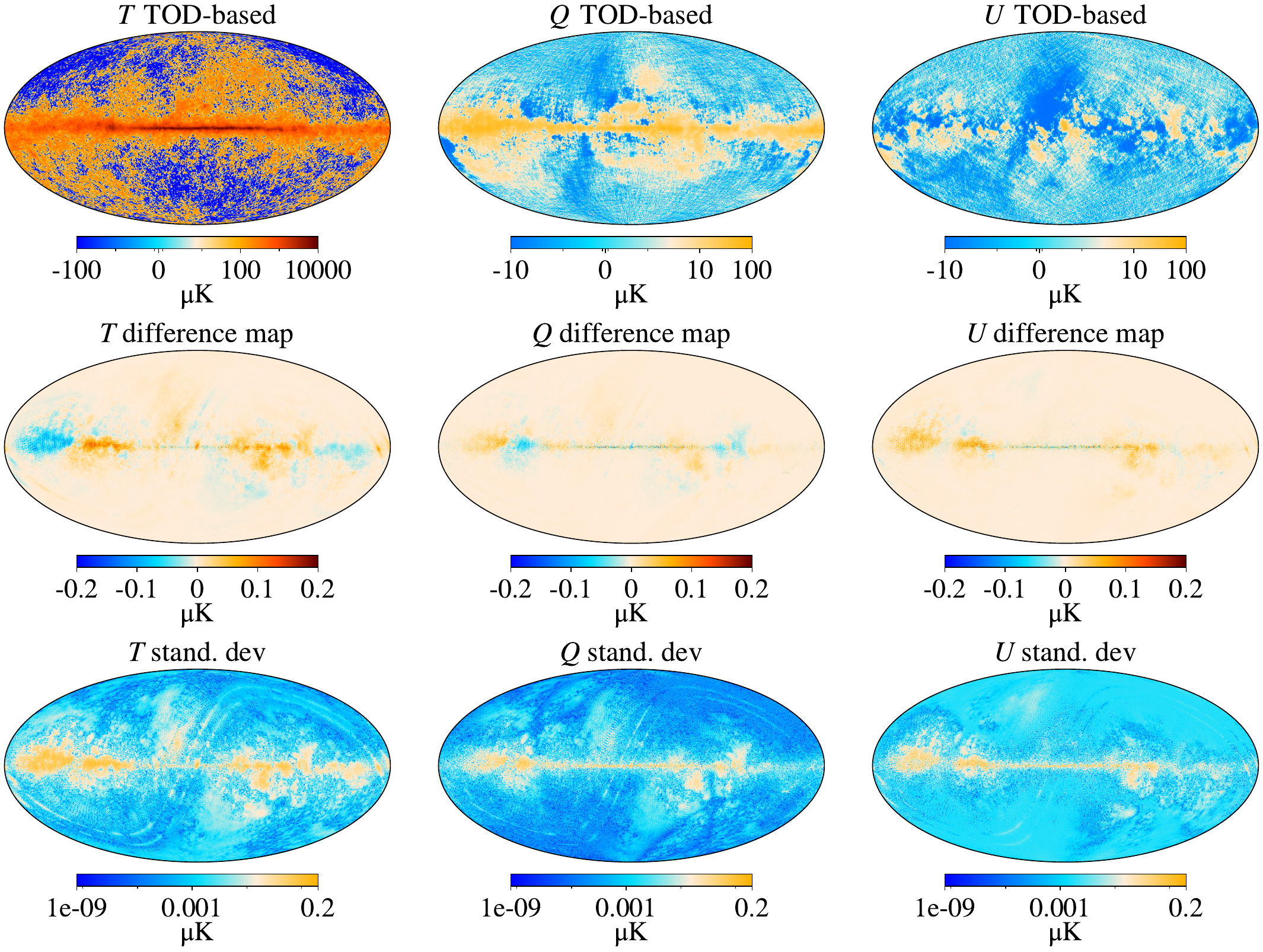}
    \caption{Validation of the analytical model against TOD-based maps. First row: TOD-based maps for channel L4-140 for temperature and polarization in the perfect calibration scenario. Second row: map differences of TOD-based and AM.
    Bottom row: the standard deviation of the map differences (see Eq.~\eqref{eqn:stand_dev_of_map_diff}) captures the uniformity of the two approaches.}
    \label{fig:validation_maps}
\end{figure}

We choose four detectors from channel L4-140 that are symmetrically placed near the center of the focal plane and use the \textit{LiteBIRD} Simulation Framework (LBS) software package\footnote{\url{https://github.com/litebird/litebird_sim}} to simulate one year of observations~\cite{LiteBIRD:2025tua}. In particular, we use the \texttt{hwp\_sys} module\footnote{\url{https://github.com/litebird/litebird_sim/tree/master/litebird_sim/hwp_sys}} to perform the bin-averaging given $\widehat{\mathcal{M}}_\textsc{hwp}=\mathcal{M}_\textsc{hwp}$ and store the output maps at $\nside=128$ \healpix\footnote{\url{http://healpix.sourceforge.net}} resolution. The maps are shown in the top row of figure \ref{fig:validation_maps}.

\begin{figure}
    \centering
    \includegraphics[width=1.\linewidth]{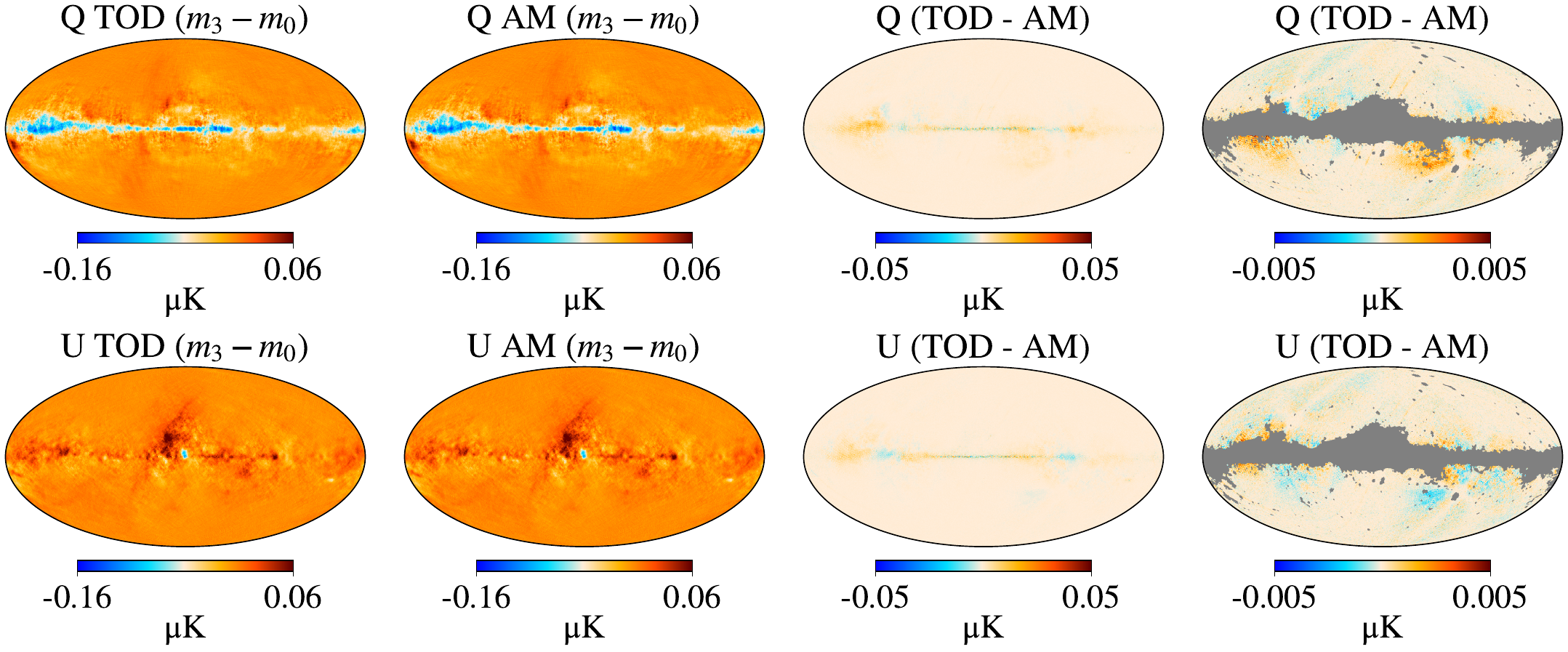}
    \caption{Difference between the miscalibrated and calibrated maps remains consistent across both simulation methods when a (phase-shift) miscalibration is introduced for channel L4-140. The comparison focuses on the cases $m=3$ and $m=0$. The first two columns show the difference maps obtained from the TOD and AM simulations, respectively. The third column displays the difference between these two, while the fourth column shows the same quantity after applying a Galactic mask. The masked difference is suppressed by an additional order of magnitude, further confirming the strong agreement between the two approaches.}
    \label{fig:validation_maps_w_systematic}
\end{figure}

To show the good agreement between the output maps returned by the TOD and AM (analytical model) simulations, we also plot the difference maps and their standard deviations (second and third row of figure \ref{fig:validation_maps}, respectively):
\begin{subequations}
\begin{align}
    \Delta\mathbf{m} &\equiv \mathbf{m}_\text{TOD} - \mathbf{m}_\text{AM}\,,\\
    \boldsymbol{\sigma} &\equiv \sqrt{\Delta\mathbf{m} - \braket{\Delta\mathbf{m}}_\mathrm{pix}}\,,\label{eqn:stand_dev_of_map_diff}
\end{align}
\end{subequations}
where the angled brackets denote the average over all the pixels.

Although some features are noticeable near the Galactic plane, where the signal intensity is higher, these are inconsequential, as shown by the uniformity of the standard deviation maps in the third row. Note that with the logarithmic color scale, faint scan‑aligned striping becomes visible outside the Galactic plane. The effect is strongly exaggerated by the logarithmic stretch, which is why the difference maps (second row) remain featureless at the same locations and the standard‑deviation panels appear closer to uniformity at high latitude. Furthermore, this particular comparison uses only four detectors; with the full focal plane, cross‑linking improves significantly and the striping is expected to largely diminish. For conservatism, spectra are still computed on masks that exclude the plane (see Sec.~\ref{subsec: spectra and likelihood}).

Furthermore, we verify that the introduction of systematic effects does not introduce inconsistencies between the two approaches. In the previous test, we assessed whether the discrepancy between the TOD and the AM remains sufficiently small by comparing the same miscalibrated map across both approaches, ensuring that $\text{TOD}_{m^\prime} - \text{AM}_{m^\prime} \rightarrow0$ for $m^\prime=0,1,2,3,4,5$. In contrast, figure~\ref{fig:validation_maps_w_systematic} extends this validation by examining whether the impact of miscalibration remains consistent between the two methods. Specifically, we test whether the difference between the miscalibrated ($m=3$ for phase-shift in this case) and perfectly calibrated maps is comparable for both approaches. The consequence of the systematic miscalibration is to introduce a distortion in the maps, and this effect is shown to be consistently captured by both methods.
\begin{figure}
    \centering
    \includegraphics[width=1.\linewidth]{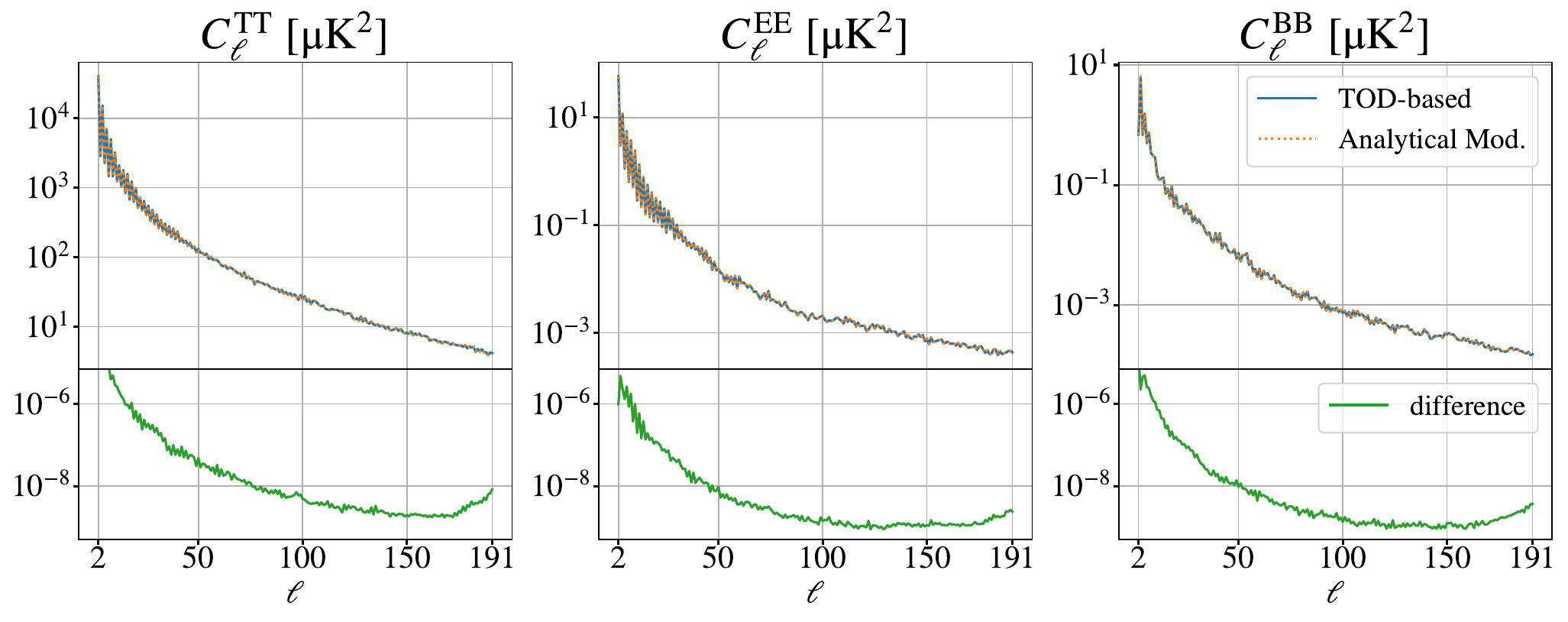}
    \caption{Power spectra for channel L4-140 maps. The upper panels show the $C_\ell$ signals from both TOD-based and analytical-model simulations of figure~\ref{fig:validation_maps}. The lower panels display the power spectra of the map differences.}
    \label{fig:validation_cls}
\end{figure}
\par
Figure~\ref{fig:validation_cls} provides further validation through the power spectra comparison. In the upper panels, we show the power spectra from maps generated by the TOD-based simulation and the analytical model of figure~\ref{fig:validation_maps}. They overlap completely, confirming the consistency between the two methods. The bottom panels show the power spectrum of the map differences, which remains nearly constant across all autocorrelated fields, further confirming the reliability of the method.

\bibliographystyle{JHEP}
\bibliography{bibliography}

\end{document}